\documentclass[fleqn,usenatbib]{mnras}

\usepackage{newtxtext,newtxmath}

\usepackage[T1]{fontenc}
\usepackage{ae,aecompl}

\usepackage{graphicx}	
\usepackage{amsmath}	
\usepackage{amssymb}	
\usepackage[bottom, hang, flushmargin]{footmisc}
\usepackage{threeparttable}
\usepackage{xspace}
\usepackage{adjustbox}

\newcommand{\aox}{\,$\alpha_{\rm{ox}}$\xspace}	
\newcommand{\daox}{\,$\Delta\alpha_{\rm{ox}}$\xspace}
\newcommand{\fopt}{\,$f_{2500}$\xspace}
\newcommand{\civ}{\,\ion{C}{iv}\xspace}
\newcommand{\civew}{\,\mbox{\ion{C}{iv} EW}\xspace}
\newcommand{\mgii}{\ion{Mg}{ii}\xspace}
\newcommand{\mgiiew}{\mbox{\ion{Mg}{ii} EW}\xspace}
\newcommand{\Chandra}{\,\emph{Chandra}\xspace}
\newcommand{\angstrom}{\text{\normalfont\mbox{\AA}}\xspace}

\newcommand*{\thead}[1]{\multicolumn{1}{c}{ #1}}

\title[Broad line--X-ray correlations]{The correlations between optical/UV broad lines and X-ray emission for a large sample of quasars}

\author[J. D. Timlin et al.]{
John D. Timlin III,$^{1 \thanks{E-mail: jxt811@psu.edu}}$
W. N. Brandt,$^{1,2,3}$
Q. Ni,$^{1}$
B. Luo,$^{4,5}$
Xingting~Pu,$^{4,6}$
\newauthor
D. P. Schneider,$^{1,2}$
M. Vivek,$^{1}$
and
W. Yi$^{1,7,8}$
\\
$^{1}$Department of Astronomy \& Astrophysics, 525 Davey Lab, The Pennsylvania State University, University Park, PA 16802, USA \\
$^{2}$Institute for Gravitation and the Cosmos, The Pennsylvania State University, University Park, PA 16802, USA \\
$^{3}$Department of Physics, 104 Davey Lab, The Pennsylvania State University, University Park, PA 16802, USA\\
$^{4}$School of Astronomy and Space Science, Nanjing University, Nanjing, Jiangsu 210093, China\\
$^{5}$Key Laboratory of Modern Astronomy and Astrophysics (Nanjing University), Ministry of Education, Nanjing, Jiangsu 210093, China\\
$^{6}$College of Science, Nanjing Forestry University, Nanjing, Jiangsu 210037, China\\
$^{7}$Yunnan Observatories, Kunming, 650216, Peoples Republic of China\\
$^{8}$Key Laboratory for the Structure and Evolution of Celestial Objects, Chinese Academy of Sciences, Kunming 650216, Peoples Republic of China\\
}

\date{Accepted 2019 December 4. Received 2019 December 3; in original form 2019 October 2.}

\pubyear{2019}

\begin{document}
\label{firstpage}
\pagerange{\pageref{firstpage}--\pageref{lastpage}}
\maketitle

\begin{abstract}
We present \Chandra observations of 2106 radio-quiet quasars in the redshift range \mbox{$1.7 \leq z \leq 2.7$} from the Sloan Digital Sky Survey (SDSS), through data release fourteen (DR14), that do not contain broad absorption lines (BAL) in their rest-frame UV spectra. This sample adds over a decade worth of SDSS and \Chandra observations to our previously published sample of 139 quasars from SDSS DR5 which is still used to correlate X-ray and optical/UV emission in typical quasars. We fit the SDSS spectra for 753 of the quasars in our sample that have high-quality (exposure time $\geq$ 10 ks and off-axis observation angle $\leq 10\arcmin$) X-ray observations, and analyze their X-ray-to-optical SED properties (\aox and \daox) with respect to the measured CIV and MgII emission-line rest-frame equivalent width (EW) and the CIV emission-line blueshift. We find significant correlations (at the $\geq 99.99$\% level) between \aox and these emission-line parameters, as well as between \daox and \civew. Slight correlations are found between \daox and \civ blueshift, \mgiiew, and the \civew to \mgiiew ratio. The best-fit trend in each parameter space is used to compare the X-ray weakness (\daox) and optical/UV emission properties of typical quasars and weak-line quasars (WLQs). The WLQs typically exhibit weaker X-ray emission than predicted by the typical quasar relationships. The best-fit relationships for our typical quasars are consistent with predictions from the disk-wind quasar model. The behavior of the WLQs compared to our typical quasars can be explained by an X-ray ``shielding'' model.

\end{abstract}

\begin{keywords}
galaxies: active -- quasars: general -- quasars: emission lines -- X-rays: general
\end{keywords}



\section{Introduction}

It is widely accepted that every massive galaxy has undergone at least one quasar phase throughout its lifetime (e.g.\ \citealt{Soltan1982, Richstone1998, Kormendy2013}) in which the central super-massive black hole (SMBH) is actively accreting gas from a surrounding accretion disk. Accretion occurs when the gas in the disk loses angular momentum through viscous transfer, causing the disk to heat and emit optical and ultraviolet (UV) radiation (e.g.\ \citealt{Salpeter1964, Lynden-Bell1969,Rees1984}). Quasars also emit highly ionizing X-ray radiation which originates from a central ``corona'' surrounding the SMBH. The nature of this corona is still uncertain; however, correlations between the optical/UV and X-ray emission in quasars have been used as a diagnostic to compare the coronal region and the accretion disk \mbox{(e.g.\ \citealt{Steffen2006, Just2007, Lusso2017})}. 

Furthermore, the ionizing X-ray radiation has been linked to the emission-line properties in rest-frame optical/UV spectra \mbox{(e.g.\ \citealt{Wilkes1987, Murray1995, Leighly2004})}. Broad emission lines are a distinctive feature in quasar spectra, and are attributed to the large orbital velocities of ionized gas around the SMBH (e.g.\ \citealt{Peterson1993,Netzer2015}). Quasar spectra are host to both high-ionization emission lines (such as \mbox{\ion{C}{iv}} and \mbox{\ion{Si}{iv}}) as well as low-ionization emission lines (e.g.\ \mbox{\ion{H}{$\beta$}} and \mbox{\ion{Mg}{ii}}). Low-ionization emission-lines are observed to have more symmetric profiles and are typically good indicators of the quasar redshift, whereas high-ionization lines, particularly the \civ emission-line, tend to be asymmetric and can exhibit a large range of outflow velocities (typically between 0--1500 km s$^{-1}$; e.g.\ \citealt{Gaskell1982, Wilkes1984,Shen2016}) in addition to a virialized component. 

Such behavior has often been explained by the widely-utilized disk-wind quasar model (e.g.\ \citealt{Murray1995,Chiang1996, Elvis2000, Proga2000}) which splits the quasar broad emission-line region (BELR) into a disk and wind component. High-ionization lines are often associated with the wind component (particularly for high-luminosity quasars), which is exposed to the full ionizing continuum of the quasar, whereas the low-ionization lines are considered to be formed in the accretion disk and are subjected to a continuum that has been modified by the wind component (e.g.\ \citealt{Leighly2004,Richards2011}). The formation and velocity of the wind is dependent on the shape of the spectral energy distribution (SED), in particular the strength of the ionizing radiation that interacts with the BELR. As the number of ionizing photons in the BELR increases, the gas in the wind region becomes more ionized, causing the abundance of high-ionization lines to increase but their observed outflow velocity to decrease \citep{Leighly2004, Leighly2007, Richards2011}. This proposal has been tested in numerous investigations, typically by correlating the X-ray-to-optical spectral slope, \aox\footnote{\aox $= 0.3838 \times {\rm{log}}_{10}(f_{2\rm\ keV}/f_{2500})$, where $f_{2 \rm\ keV}$ and $f_{2500}$ are the flux densities at rest-frame 2 keV and 2500 \angstrom, respectively.}, and \civ emission-line properties (e.g.\ \citealt{Green1998, Gibson2008, Wu2009, Wu2011, Richards2011, Kruczek2011}).

Integral to many of the investigations into the \aox--\civ correlations was the data sample compiled in \citet{Gibson2008}. This work compared the optical/UV properties of spectroscopically-confirmed quasars from the Sloan Digital Sky Survey (SDSS; \citealt{York2000}) data release five (DR5; \citealt{Adelman-McCarthy2007}) with their X-ray counterparts, from {\emph{Chandra}} or {\emph{XMM-Newton}}, in order to constrain physical models that relate X-ray and UV emission. In total, they identified 536 quasars with overlapping X-ray coverage in the redshift range $1.7 \leq z \leq 2.7$ (this choice of redshift range is described in Section \ref{sec:sample_selection}) to use in their investigation. Their full sample, however, contained many X-ray non-detections, particularly due to the larger limiting flux value of {\emph{XMM-Newton}}, so they reduced the sample to quasars that were only selected homogeneously in the \Chandra fields (their Sample B). Sample B of \citet{Gibson2008} contains 139 quasars that do not have broad absorption line (BAL) features and are radio-quiet (hereafter, ``typical quasars''). Since this sample is homogeneously selected, correlations between X-ray and optical/UV properties are less likely to be affected by unknown selection effects (e.g.\ due to observation flux limits). Sample B, however, contains too few quasars to compute correlations with high confidence, and therefore \citet{Gibson2008} used the full sample of quasars to fit general X-ray-optical/UV correlations, despite the large number of X-ray non-detections. Moreover, the quasars selected in SDSS DR5 are typically at the bright end of the quasar luminosity function, thus perhaps not sampling a diverse set of quasar properties. We seek to increase the sample size of quasars with unbiased X-ray measurements to better understand the nature of the X-ray corona and BELR in typical quasars by correlating their X-ray and optical/UV emission properties.

In addition to finding correlations among typical quasars, Sample B has also been used to compare properties of the typical quasars with a population of quasars that have weak \civ emission lines. These weak-line quasars (WLQs) are a unique subset of the radio-quiet quasar population, and are classified by their notably small \ion{C}{iv} $\lambda 1549$ equivalent widths (EWs), with \mbox{\civew $\leq 15$ \angstrom} (unless specified otherwise, EW refers to rest-frame equivalent width). In many cases, the \civ line profile in WLQs also exhibits extremely large blueshifts (as high as \mbox{$\approx$10,000 km s$^{-1}$}; e.g.\ \citealt{Wu2011, Wu2012, Luo2013, Luo2015, Plotkin2015, Ni2018}), where the peak of the \civ emission line is shifted to a smaller wavelength in the rest frame than the vacuum wavelength (\mbox{$\lambda$ = 1549.06 \AA }), which is indicative of an outflow.  Moreover, X-ray observations of WLQs have revealed that approximately half of the population has \mbox{$\approx$20--70} times weaker \mbox{X-ray} emission than expected,\footnote{The expected X-ray weakness is derived from the monochromatic rest-frame 2500 \angstrom luminosity, $L_{2500}$, of the WLQ and the relationship between \aox and $L_{2500}$ for the typical quasars (e.g.\ \citealt{Just2007}).} and may have significant X-ray absorption along the line-of-sight \citep{Luo2015}. The other half of the WLQ population appears to be X-ray normal, with a measured average X-ray power-law spectral index indicative of high Eddington-ratio accretion \citep{Luo2015, Marlar2018}. 

Such extreme properties are not generally observed among typical quasars, resulting in an additional ``shielding'' model to explain the observed behavior of the WLQs in the context of the overall quasar population \citep{Luo2015, Ni2018}. This model asserts that the inner accretion disk in quasars with high Eddington ratios becomes geometrically thick, and blocks the ionizing X-ray and extreme-UV continuum from reaching the broad emission-line region. Without these ionizing photons in the BELR, the EW (which we use as a proxy for abundance) of \civ ions in the BELR decreases, and the emission-line blueshift increases, which indicates the presence of a high velocity outflow. The geometrically thick disk is also responsible for obscuring the X-ray emission along the line-of-sight when the inclination angle of the system is large. While most WLQ investigations were performed using small samples \citep{Wu2011, Wu2012}, \citet{Luo2015} used a sample of 51 WLQs to test the ``shielding" model. Adding to this work, \citet{Ni2018} combined the previously known WLQs, as well as new observations, to create a sample of 63 WLQs to examine possible correlations between \mbox{X-ray} and optical/UV emission, also in the context of the ``shielding'' model. Both investigations compared the X-ray and \civ properties of the WLQs to the typical quasars in Sample B of \citet{Gibson2008} to test for relationships between WLQs and typical quasars, and to demonstrate the extreme X-ray and optical/UV properties of WLQs.

In this analysis, we assemble a large, unbiased sample of typical quasars, akin to Sample B, in order to constrain better correlations between the X-ray and optical/UV broad emission-line properties, and to compare the relationships derived for typical quasars with measurements for the large WLQ sample in \citet{Ni2018}. Since 2008, the SDSS has discovered and published nearly seven times the number of spectroscopically-confirmed quasars than are in SDSS DR5.\footnote{And approximately nine times the number within $z =$ 1.7--2.7, the redshift range of interest.} There has also been $\approx$10 additional years of high-quality \Chandra observations, substantially expanding the solid angle of sensitive X-ray coverage across the sky. Including these new data greatly improves the constraints on the X-ray and optical/UV emission-line correlations and, in turn, allows proper comparison of the typical quasars with the WLQs. Furthermore, the larger data set samples more diverse typical quasars (e.g.\ a wider range of luminosity, Eddington ratio, shape of the spectral energy distribution, etc.), granting access to regions in the parameter spaces not covered by Sample B. 

This paper is organized as follows: Section 2 describes the data sets used to assemble a sample of typical quasars for this investigation, and presents the method used to find X-ray counterparts. The data-analysis techniques used to reduce and analyze the X-ray data and the software used to fit the optical/UV broad emission lines are discussed in Section 3. Section 4 correlates the relative X-ray strength with various measured emission-line properties. Finally, the implications of these results for both the typical quasars and the WLQs are reviewed in Section 5, and our conclusions are summarized in Section 6. Throughout this work, we adopt a flat $\Lambda$-CDM cosmology with \mbox{$H_{0}$ = 70 km s$^{-1}$ Mpc$^{-1}$}, \mbox{$\Omega_M$ = 0.3}, and \mbox{$\Omega_{\Lambda}$ = 0.7}, and we utilize the \Chandra Interactive Analysis of Observations (CIAO; \citealt{Fruscione2006}) version 4.10\footnote{\url{http://cxc.harvard.edu/ciao/releasenotes/ciao_4.10_release.html}} software and CALDB version 4.8.3.\footnote{\url{http://cxc.harvard.edu/caldb/}} 


\section{Sample Selection}\label{sec:sample_selection}

To increase the sample size of the unbiased Sample B of \citet{Gibson2008}, we select typical quasars from the seventh, twelfth, and fourteenth data releases of the SDSS quasar catalog (DR7Q; \citealt{Schneider2010}; DR12Q; \citealt{Paris2017}; DR14Q; \citealt{Paris2018}) that span a redshift range of 1.7$\leq z \leq$ 2.7. As in \citet{Gibson2008}, the lower redshift limit grants sufficient spectral coverage to identify and remove highly blueshifted \civ BALs; the upper redshift limit ensures that rest-frame 2500 \mbox{\AA} has good photometric coverage in the SDSS camera for each quasar. Below we provide a brief description of each SDSS catalog and the restrictions we make on the optical data, as well as the method used to find serendipitously observed counterparts in \Chandra to generate a large, unbiased sample of quasars.

\subsection{Optical data}

The SDSS DR7Q catalog contains all $105,783$ spectroscopically-confirmed quasars observed in the first two SDSS projects (\mbox{SDSS-I/II}; \citealt{York2000}) over an area of $\approx9380$ deg$^2$ across the sky. These quasars span a wide redshift range ($0.065\leq z \leq 5.46$) with observational flux-density limits for targeted low- and high-redshift quasars of $i=19.1$ and $i=20.2$, respectively. \citet{Shen2011} measured the continuum and emission-line properties of each quasar spectrum, and reported their measurements along with the photometric properties from the DR7 catalog. Also included in that work is a flag indicating the presence of a BAL and a parameter measuring the radio-loudness of each quasar, which was computed using flux-densities from the Faint Images of the Radio Sky at Twenty centimeters (FIRST; \citealt{Becker1995}) survey reported in DR7Q. \citet{Shen2011} also reported the improved redshift measurements from \citet{HW2010}, which tend to be more reliable than the redshifts reported by the SDSS pipeline. To generate an unbiased sample akin to Sample B, we used these improved redshifts to remove quasars outside of the redshift range 1.7 $\leq z \leq$ 2.7. We  then removed quasars that are flagged as BALs ({\tt{BAL\_FLAG}=}0) and quasars that have extremely red intrinsic colors ($\Delta(g-i) \leq 0.45$; see \citealt{Richards2003}), which eliminated potentially X-ray absorbed quasars. These criteria are consistent with those imposed on the WLQs \citep{Ni2018} so that we can compare the two samples. 

To the DR7 subset, we added the quasars observed in the \mbox{SDSS-III} campaign \citep{Eisenstein2011}, all of which are reported in SDSS DR12Q \citep{Paris2017}. This catalog contains $297,301$ spectroscopically-confirmed quasars, and spans a large range in redshift ($0.04<z<6.44$). SDSS-III was designed to observe a large number of quasars at moderate redshifts ($2.15<z<3.5$), where the first two SDSS cycles were significantly incomplete. The redshift distribution peaks around $z\approx2.5$, which is within the targeted redshift range of our investigation, thus greatly increasing the sample size. Additionally, DR12Q reports the measured properties of the \civ emission-line, the balnicity index \citep{Weymann1991}, a BAL visual inspection flag which we use to remove BALs, and quasars with radio detections in FIRST. We imposed similar cuts to DR12Q as we made in DR7Q, keeping only non-BAL ({\tt{BI\_CIV}=0} {\tt{or}} {\tt{BAL\_FLAG\_VI}=}0) quasars within our targeted redshift range (where we adopt the visual-inspection redshifts reported in DR12Q) that did not have very red intrinsic colors (again, $\Delta(g-i) \leq 0.45$). Finally, for DR12Q, we impose a limit on the signal-to-noise ratio (SNR) of the continuum emission at 1700 \angstrom ({\tt{SNR\_1700}>}1) as a first order cut to remove quasars that have a low spectral quality, and thus have highly uncertain emission-line measurements.

Lastly, we compile a list of quasars from the SDSS DR14Q \citep{Paris2018}, which reprocessed the spectra from DR12Q, and contains an additional $144,046$ new quasars discovered in \mbox{SDSS-IV} \citep{Dawson2016} with 2044 deg$^2$ of additional coverage on the sky. While DR14Q is a compilation of SDSS-I/II/III/IV, the peak of the redshift distribution of the quasars in SDSS-IV is \mbox{$z\approx1.7$} which, again, greatly increases the sample of quasars that can be used to find X-ray counterparts. Since DR14Q is such a large catalog, however, only a few quasar spectra were visually inspected to obtain a redshift;\footnote{We visually inspect all of the spectra for quasars in DR14Q that are in our final catalog.} many of the redshifts in the catalog were obtained by performing a principle component analysis (PCA) on the spectrum (see \citealt{Paris2012}). In our investigation, we use the visually inspected redshifts when available and the PCA redshift otherwise (with {\tt{Z\_PCA\_ERR}=0} ). DR14Q also reports measurements of the balnicity index and the photometric properties of the quasars, which allowed us to impose the same constraints on the data as before.

The quasars in DR7Q, DR12Q, and DR14Q that satisfied our initial cuts were combined to form a catalog of SDSS quasars ($\approx97,000$ quasars) that we used to search the \Chandra database for serendipitous observations. In cases where sources overlapped between catalogs, we preferentially retained the earliest detection of the quasar (e.g.\ for objects observed in all three catalogs, we retained the DR7Q information and spectral measurement). We preferentially kept the DR7Q observations because they have both more reliable flux calibrations and available high-quality spectral measurements from \citet{Shen2011} which were used to test our spectral measurements (see Section~\ref{sec:specfit}). Furthermore, we computed the value of the radio-loudness parameter (or an upper limit) for each of our quasars.\footnote{The radio-loudness parameter is defined as the ratio between the 6 cm flux density and the 2500 \angstrom flux density: $R = f_{6\rm{cm}}/f_{2500}$ \citep{Richards2011}.} Radio-loud quasars tend to have additional X-ray emission associated with their jets (as opposed to just coronal emission), so restricting to quasars that are not radio-loud removes biases due to different regions in the quasar emitting X-ray radiation (e.g.\ \citealt{Miller2011}). The radio-loudness parameter is recorded in the \citet{Shen2011} quasar catalog; however, we computed the parameter (or an upper limit) for the quasars from DR12Q and DR14Q. 

To compute the $R$-parameter, we first estimated \fopt, following the method in \citet{Richards2006}, by converting the apparent $i$-band magnitudes (corrected for Galactic extinction; \citealt{Schlafly2011}) to absolute magnitudes. The absolute magnitudes were then K-corrected to $z=2$ ($M_i(z=2)$) using the K-corrections in \citet{Richards2006}. These corrections assume a constant spectral index of $\alpha_{\nu}=-0.5$ \citep{Vandenberk2001} and subtract the average quasar emission-line signatures that enter the $i$-band at this redshift. Then, using Equation 4 in \citet{Richards2006}, we convert $M_i(z=2)$ to monochromatic luminosity at 2500 \mbox{\AA}, which was finally converted\footnote{$L_{2500} = 4 \pi D_{L}^{2} f_{2500}/(1+z)$ where $D_L$ is the luminosity distance.} to \fopt. The rest-frame 6 cm flux density was computed using the 20 cm flux density reported in the catalogs for objects that are detected in FIRST \citep{Becker1995}, and assuming a radio spectral index of  $\alpha_{\nu}=-0.5$. Many quasars, particularly for DR12 and DR14, were not detected in FIRST; therefore, we derived $2\sigma$ upper limits on the 20 cm radio flux using $0.25 + 2\sigma_{\rm rms}$ mJy, where $\sigma_{\rm rms}$ is the RMS flux at the source position and 0.25 mJy is the CLEAN bias correction \citep{White1997}. Some of the quasars in our sample were not covered by FIRST so we searched the NRAO VLA Sky Survey (NVSS; \citealt{Condon1998}) database\footnote{See \url{https://www.cv.nrao.edu/nvss/NVSSPoint.shtml}} instead to obtain radio fluxes and limits. Quasars are typically considered radio-loud when $R > 10$; therefore, we removed such quasars from our sample if they were detected in radio wavelengths. Objects with upper-limits on the $R$-parameter were retained, and we will discuss them further in Section \ref{sec:Xdata_redux}. We use the \fopt values calculated here throughout the rest of this investigation since the spectral flux density calibration in DR12Q and DR14Q has large uncertainties, particularly at the red end of the spectrum \citep{Paris2012,Margala2016}. We find that our \fopt values are largely consistent with the values in \citet{Shen2011}, which were measured from the SDSS DR7 spectra. After imposing all of the conditions above, we investigated which quasars had overlapping X-ray coverage.

\subsection{Finding \Chandra counterparts}
Most of our quasars are optically fainter than those from the \citet{Gibson2008} sample, making the high-sensitivity observations from \Chandra ideal for this investigation. Since these quasars tend to be faint, we elected to perform forced-photometry of the SDSS quasars using the \Chandra events files rather than match these quasars to the \Chandra Source Catalog (CSC; \citealt{Evans2010}).\footnote{See \url{http://cxc.harvard.edu/csc/}} Performing the analysis in this manner allows us to examine these quasars to a greater sensitivity than would be provided by a simple match to the CSC catalog, and enables us to generate constraints on quasars that are not detected.

Following the method in \citet{Gibson2008} to generate their unbiased Sample B, we searched for serendipitous observations of our SDSS quasar sample in the \Chandra archive. Restricting our analysis to only the serendipitously observed sources removes any potential biases that might occur in cases where \Chandra targeted quasars with atypical properties.\footnote{\citet{Gibson2008} found that removing targeted sources had little effect on their results; however we conservatively choose to remove as many biases as possible.} To identify serendipitous sources, we first compared the positions of each of our quasars to the Multi-Order Coverage (MOC\footnote{\url{http://cxc.cfa.harvard.edu/cda/cda_moc.html}}) map which approximates the regions of the public \Chandra observation footprints. Approximately 3\% of our SDSS quasar sample is contained within the \Chandra footprint, so this step quickly removes the large fraction of our objects which did not have X-ray coverage. 

The {\tt{find\_chandra\_obsid}} tool provided in \Chandra Interactive Analysis of Observations (CIAO; \citealt{Fruscione2006}) was used to find the observation ID associated with the given position of the quasar; we searched only for detections with the ACIS instrument \citep{Garmire2003} where no gratings were used in the observation. On occasions when more than one observation ID was found, we flagged the object as having multiple observations, and retained all of the information for each observation ID. The MOC maps approximated the coverage of the footprint, so it did not guarantee that the position of the quasar is on an ACIS CCD, i.e.,\ the quasar could lie on the edge of the detector. We removed observations too close to an edge by requiring that the position of the quasar be more than 40 pixels ($\approx20\arcsec$) from the edge of the image (e.g.\ \citealt{Gibson2008}). This allows retention of only high-quality observations that are not artificially biased by bad measurements for the quasar (or background) near an edge of the detector. Finally, for objects with multiple \Chandra observations, we retained only the ID with the smallest off-axis angle\footnote{The distance between the \Chandra aim-point and the location of the quasar measured in arcminutes.} and the largest exposure time, which ensured that we obtain the most sensitive \Chandra observation for each of our target quasars. 

\begin{figure}
	\includegraphics[width=\columnwidth]{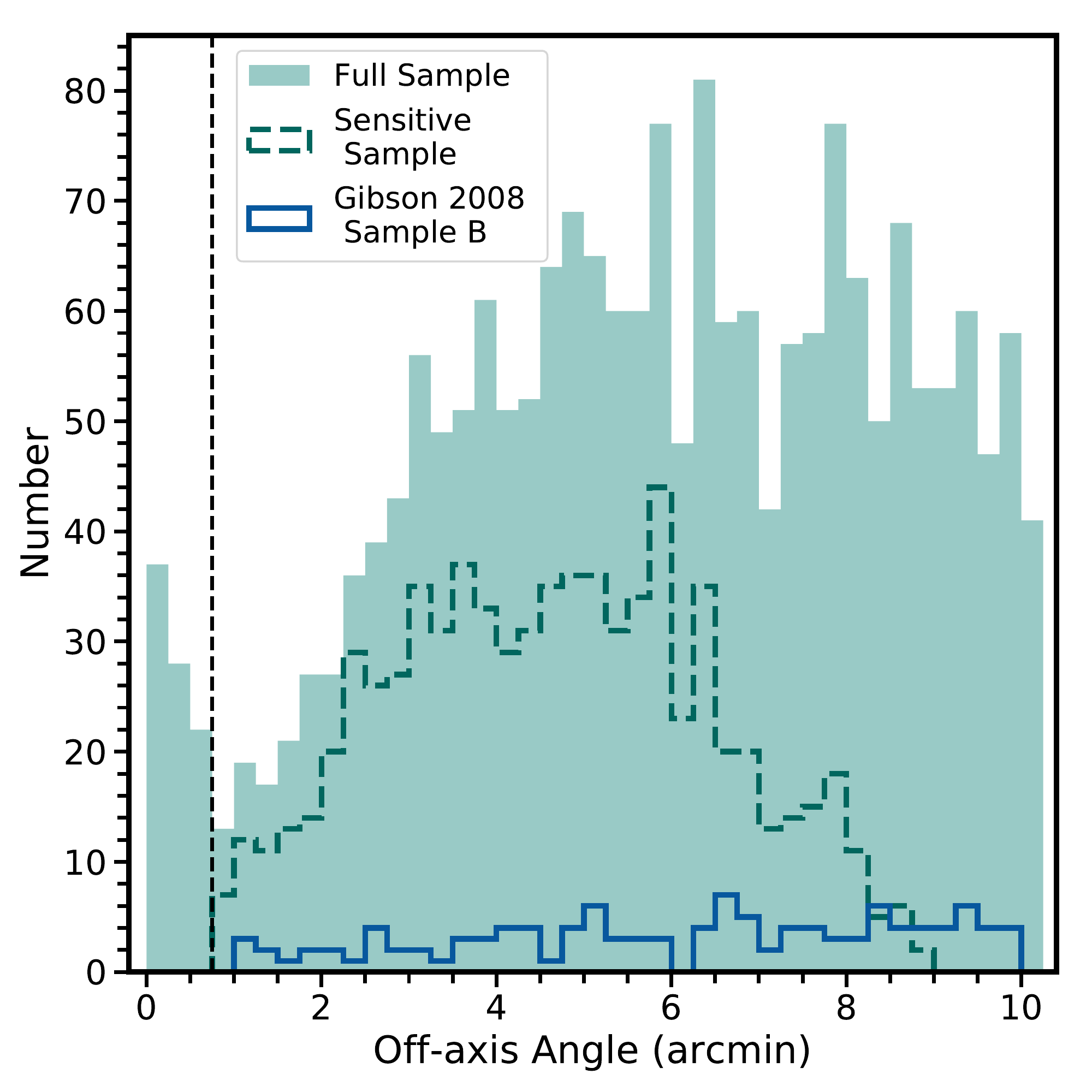}
    \caption{Number of typical SDSS quasars with suitable serendipitous \Chandra coverage in our Full data set (light green histogram; 2106 sources) as a function of the off-axis angle from the \Chandra aim-point. In this analysis, we exclude objects that lie within 0.75\arcmin \ (vertical dashed line) of the aim-point to generate an unbiased sample of serendipitously observed quasars (2019 quasars). The Sensitive sample from Section \ref{sec:Xdata_redux} (dark green dashed line; 753 sources) contains an unbiased subsample of quasars from the Full data set. The Sample B quasars (solid blue line; 139 sources) are depicted as a comparison between the sizes of these samples.}
    \label{fig:off-axis_hist}
\end{figure}

In total, 2106 SDSS quasars passed the above criteria\footnote{466 quasars from DR7Q, 1144 quasars from DR12Q, 496 quasars from DR14Q.} (hereafter, the Full sample). We present the distribution of their off-axis angles in Figure \ref{fig:off-axis_hist} along with those of Sample B of \citet{Gibson2008}, where we impose a cut in off-axis angle of $0.75\arcmin$ to ensure that the \Chandra observation is indeed a serendipitous observation (2019 objects remain after the application of this constraint).\footnote{No quasars in our sample that were observed on-axis also had serendipitous observations within 10\arcmin.}  Figure \ref{fig:off-axis_hist} also illustrates the vast increase in the number of confirmed quasars that have been serendipitously observed since the investigation of \citet{Gibson2008} which can be used to assess the X-ray and optical/UV broad line properties of typical quasars. Table \ref{tab:Samples} reports the basic properties for our Full sample, as well as the subsamples which we will explain in greater detail below. 

\begin{figure}
	\includegraphics[width=\columnwidth]{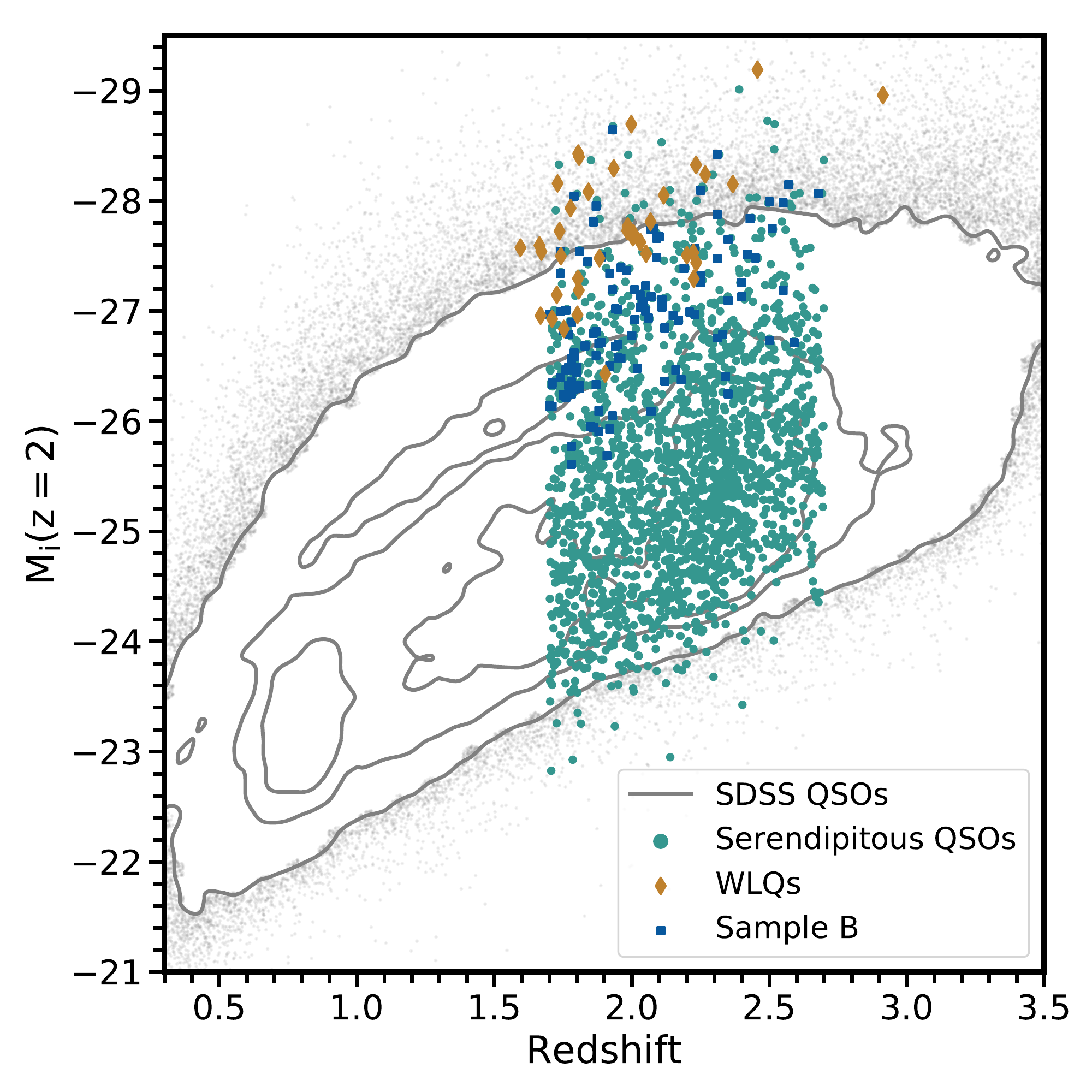}
    \caption{Absolute $i-$band magnitude (K-corrected to $z=2$; see \citealt{Richards2006}) as a function of redshift for the 2106 SDSS quasars that are in the field-of-view of a \Chandra observation (green points). The redshift range of our sample ($1.7 \leq z \leq 2.7$) has been chosen to be consistent with the redshifts of the WLQs. Depicted in the grey contours is the distribution of $M_{i}$($z=2$) and redshift for all non-BAL quasars in SDSS where each contour encloses 35, 68, and 95 percent of the quasars. For comparison, we also display the WLQs (brown diamonds) from \citet{Ni2018}, which are selected from bright SDSS quasars, and the Sample B quasars (blue squares) from \citet{Gibson2008}. Our Full sample of quasars spans a significantly larger range of luminosity than the WLQs and Sample B, and is representative of the entire sample of SDSS quasars.}
    \label{fig:Mi_z}
\end{figure}

Along with the increased sample size, our Full quasar sample also spans a wide range in luminosity. Figure \ref{fig:Mi_z} depicts the absolute $i-$band magnitude of our Full quasar sample compared to Sample B, all SDSS quasars (through DR14), and the sample of WLQs from \citet{Ni2018}. Again we measure absolute magnitudes that have been K-corrected to $z=2$ using the K-corrections in \citet{Richards2006}. Figure \ref{fig:Mi_z} illustrates that our Full quasar sample spans the luminosity range representative of all SDSS quasars, and thus our sample also allows investigation of the relationship between X-ray strength and luminosity (e.g.\ \citealt{Steffen2006, Just2007, Lusso2017}) over a wide luminosity range. The WLQs tend to have brighter absolute magnitudes; however, this trait is a selection effect since bright WLQs were selected for \mbox{X-ray} observation to allow for economical snapshot observations with \Chandra.

\section{X-ray and Optical Data Analysis}\label{sec:Xopt_data}
\subsection{X-ray data reduction}\label{sec:Xdata_redux}
Our data were processed using standard CIAO tools \citep{Fruscione2006} beginning with the {\tt{chandra\_repro}} tool, which re-calibrates the existing files into a bad pixel file, an event file, and a spectrum of counts as a function of energy. This step also accounted for the ACIS observation mode ({\tt{FAINT}} or {\tt{VFAINT}}) of each observation ID using the {\tt{check\_vf\_pha}} option, ensuring that the backgrounds were correctly cleaned for observations taken in ACIS {\tt{VFAINT}} mode. After reprocessing, we employed the {\tt{deflare}} tool which removes background flares in the light curve above the 3$\sigma$ level, and adjusts the ``good time'' intervals accordingly. From this cleaned time interval, {\tt{dmcopy}} was used to split the data into a soft (0.5--2 keV) and a hard (2--7 keV) band, and generated images according to the good ASCA grades (0, 2, 3, 4, 6).

After the images were created for each band, {\tt{wavdetect}} was run on each image using the $\sqrt{2}$ wavelet scheme (1, 1.414, 2, 2.828, 4) to search for sources in the image, where the default false detection probability was set at $10^{-6}$. For each detected source, we compared the positions from {\tt{wavdetect}} with those in SDSS and, if the coordinates of a detected source in either band matched within $0.5\arcsec$ of the SDSS position, we retained the measurement from {\tt{wavdetect}}, otherwise, we adopt the SDSS position. Using this position, we extracted the raw counts within a circular aperture with a radius equal to the 90\% enclosed-energy fraction plus 5 additional pixels \citep{Gibson2008}. Defining the radius in this manner forces the radius of the aperture to increase in response to an increasing off-axis angle. This definition also motivated the removal of objects close to the edge of a field-of-view (FOV), to ensure that the source extraction region did not overlap the edge of the image. 

From the source position and aperture radius, we generated a background region in two ways. First, if the quasar position was not near any other sources detected by {\tt{wavdetect}} in either band, the background was extracted in an annulus with radii equal to the source radius plus 15 (50) pixels for the inner (outer) regions. For many of our serendipitous detections, however, this annulus overlapped with faint sources detected in either the soft or hard band. In that case, we systematically searched for a nearby, source-free location to measure the background in a circular aperture with a radius that approximately covers the same area as the annulus region (on average, the circular background region center is $\approx 70\arcsec$ from the source center). We visually inspected all of our background regions to ensure that they were not overlapping with sources or with CCD gaps. We also required the background region to lie no less than ten pixels from the edge of any CCD boundary, as defined in the FOV files. When the target of the \Chandra observation was a galaxy cluster, the background region was contaminated by the enhanced background from the cluster (which occurred in $\approx 20$ of our observations). In such cases, we placed the background region in a local, point-source-free region by hand that was representative of the source background.

Before extracting the counts, we employed the {\tt{flux\_image}} tool to generate an exposure map for both the soft and hard bands. From the exposure map, we estimate the effective exposure time, which is the exposure time corrected for vignetting, in both the source and background regions. Finally, we use the {\tt{dmextract}} tool on both the soft and hard bands to obtain the source and background counts, exposure times, and aperture areas for each quasar, and used the {\tt{specextract}} tool to create the source and background spectral files. 

\begin{figure}
	\includegraphics[width=\columnwidth]{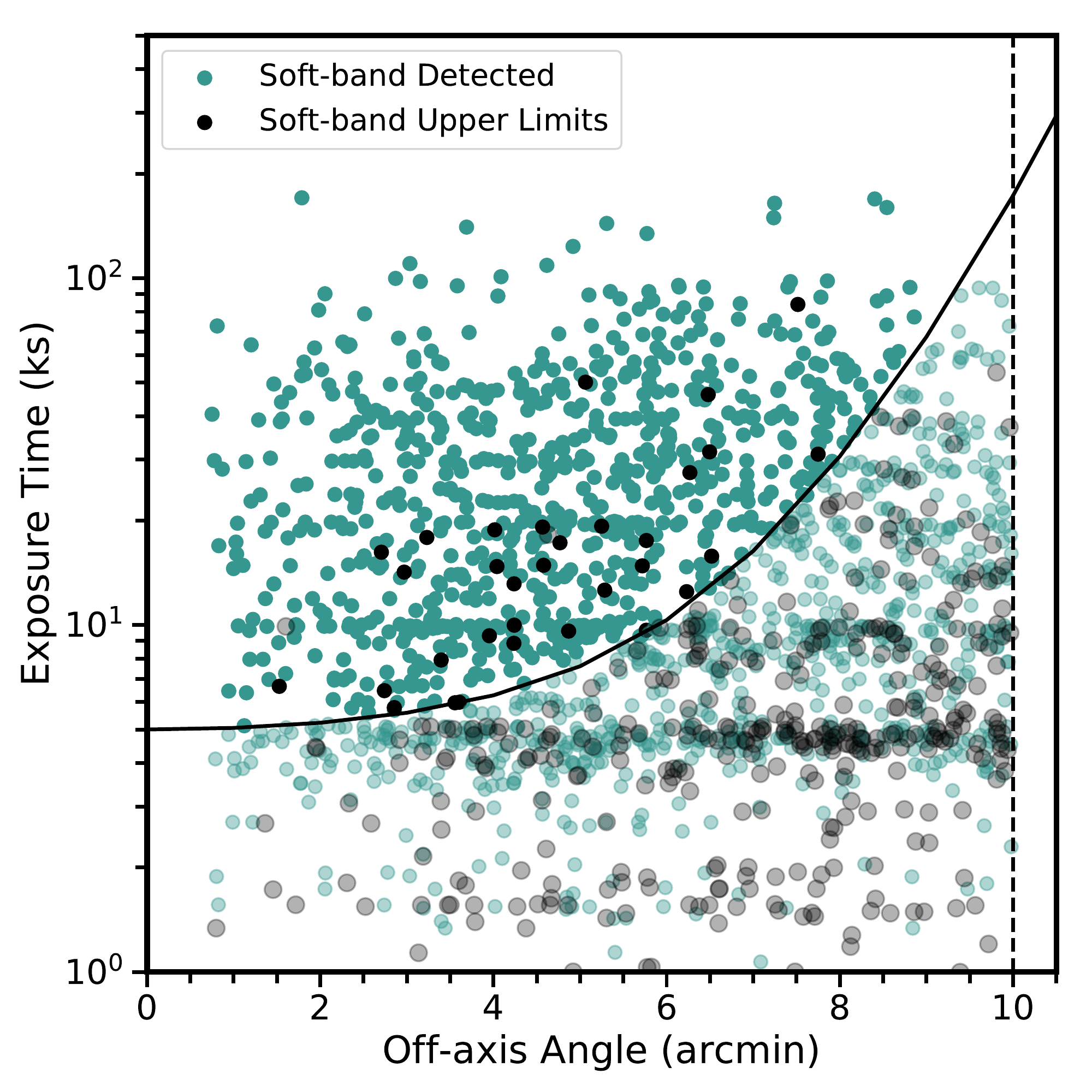}
    \caption{Effective exposure time as a function of the off-axis angle for all 2106 serendipitous sources. Green data points depict sources detected following the binomial probability threshold; the black points represent upper limits. Generally, as the off-axis angle increases, longer, more sensitive observations are necessary to obtain a detection of a serendipitous source. To generate an unbiased sample, we set a sensitivity threshold (black curve) which requires longer exposures with increasing off-axis angle, and omit objects with off-axis angles larger than 10\arcmin \ (vertical dashed line) where measurements become more uncertain even for long exposures. Our Sensitive sample of 753 quasars contains only objects located above the black curve.}
    \label{fig:exposure_offaxis}
\end{figure}

We estimated the source detection significance for the soft and hard bands using the raw counts for both the source and background. We computed the binomial no-source probability, $P_B$, to determine the probability that the source was simply detected due to random background events (\citealt{Broos2007}; \citealt{Xue2011}; \citealt{Luo2013,Luo2015}) using the formula:
\begin{equation}
P_B(X \geq S) = \sum^{N}_{X=S} \frac{N!}{X!(N - X)!}p^{X}(1 - p)^{(N - X)},
\end{equation}
where $S$ is the total source counts, $N$ is the total number of counts (source and background), and $p = 1/(1+ f_{\rm area})$ where $f_{\rm area}$ is the ratio of the background to source region area. In our analysis, we consider a source to be detected if $P_B \leq 0.01$ (2.6$\sigma$), and calculate the 1$\sigma$ errors on the net source counts following the numerical method of \citet{Lyons1991}, using the \citet{Gehrels1986} estimate of the 1$\sigma$ errors on the raw counts for the source and background. In cases where the source was not detected in an energy band, we estimated the 90\% confidence upper limit using the method of \citet{Kraft1991}. We depict the effective exposure time as a function of off-axis angle in Figure \ref{fig:exposure_offaxis} for both the detections and upper limits in our Full quasar sample.

We computed the band ratio (i.e.,\ the ratio of the hard to soft band counts) using the Bayesian Estimation of Hardness Ratio (BEHR; \citealt{Park2006}) package  for both detected sources as well as single-band non-detections, for which we can obtain limits on the band ratio. If the source is detected in the soft band but not the hard band, we consider the band ratio to be an upper limit; if the source is detected in the hard band but not the soft, the ratio is a lower limit. In cases where the source was not detected in either band, we obviously cannot place constraints on the band ratio. 

Furthermore, we estimated the effective power-law photon index, $\Gamma_{\rm eff}$, using the {\tt{modelflux}} tool in the CIAO package which converts count rates in one band into energy fluxes or count rates in a different band given the spectral response files and a spectral model, in our case a power-law modified by Galactic absorption.\footnote{This value was computed using the Leiden/Argentine/Bonn (LAB; \citealt{Kalberla2005}) measurements from the HEAsoft {\tt{Nh}} tool; see \url{https://heasarc.gsfc.nasa.gov/cgi-bin/Tools/w3nh/w3nh.pl}} We minimize the difference between the measured band ratio from BEHR and the prediction from {\tt{modelflux}} to determine $\Gamma_{\rm eff}$. As before, if the source counts are upper limits, we cannot measure the photon index, but instead adopt the value $\Gamma_{\rm eff} = 1.9$, which is the median value for typical quasars such as those in our sample (e.g.\ \citealt{Scott2011}). Using the effective photon index and the measured soft-band count rate, {\tt{modelflux}} was used to estimate the flux in the soft band and, subsequently, the rest-frame flux density at 2 keV, $f_{2\rm{keV}}$ (recall our redshift range is $1.7 \leq z \leq 2.7$, and thus our soft band always includes rest-frame 2~keV). Since we have removed reddened quasars and BAL quasars from our analysis, we expect our quasars to be largely free of substantial intrinsic X-ray absorption (e.g.\ see \citealt{Brandt2000, Scott2011}). Furthermore, the median power-law photon index ($\Gamma_{\rm eff} = 1.7$; interquartile range of 1.4--2.0) of the quasars in our sample that are detected in both bands is indicative of a quasar population without substantial intrinsic X-ray absorption; therefore, we expect the simple power-law model generally to provide an appropriate measure of the intrinsic flux of the quasars in our sample.

While we computed the aforementioned X-ray parameters for the entire Full quasar sample, we impose further constraints that remove the X-ray observations that have low sensitivity in order to generate an unbiased sample with uniformly high-quality X-ray coverage. We impose a limit on effective exposure time and off-axis angle (e.g.\ Zhu et al., in preparation), depicted by the black curve in Figure \ref{fig:exposure_offaxis}, where the quasar must lie above the empirical function:
\begin{equation}
T_{\rm eff} \geq 4.0+10^{(\theta/6.7)^2},
\end{equation}
where $T_{\rm eff}$ is the effective exposure time (in ks) at the off-axis angle, $\theta$ (in arcminutes). This restriction accounts for the decreasing sensitivity at larger off-axis angles by requiring that the observation have longer effective exposure at those positions. There are 753 quasars that remain after this cut (hereafter, the Sensitive sample; see Table \ref{tab:Samples}), all of which were visually inspected (both soft- and hard-band images) to ensure there were no issues with the images or the source/background aperture locations. As tabulated in Table \ref{tab:Samples}, this empirical restriction improves the soft-band detection fraction of our sample from 78\% in the Full catalog to $\approx$ 96\% in the Sensitive sample.\footnote{Two of the soft-band upper limits in the Sensitive sample are detected in the hard band.\label{foot:limits}} We also create a smaller sample of quasars with a higher detection fraction from the Sensitive sample by imposing an $i$-band magnitude cutoff at $i=20.2$; see Appendix~\ref{sec:appendix} for details.

Finally, we investigated whether any quasars in the Full sample or the Sensitive sample should be removed due to a large upper-limit on the radio-loudness parameter. We find that 56\% of our Full sample quasars have upper limits of $R\leq10$, the typical cutoff value between radio-loud and radio-quiet, 93\% of the quasars have $R\leq30$, and 99\% have $R\leq50$. In the Sensitive sample, which we use throughout the rest of this paper, we find 52\% of quasars have upper limits of $R\leq10$, 93\% have $R\leq30$, and 99\% have $R\leq50$. Since the $2\sigma$ upper limits on the radio-loudness parameter are only moderately radio-strong compared to the optical emission, and since only $\approx$10\% of quasars are observed to be radio-loud generally, we are confident that the quasars in our sample are radio-quiet, or at most mildly radio-intermediate, and thus our results will not be substantially affected by the non-coronal X-ray emission observed in radio-loud quasars (e.g.\ \citealt{Miller2011}). We test this assertion by removing very optically faint ($i < 21.3$) quasars, which tend to have the largest radio upper limits, from our sample. This cut does not significantly change the results presented below.

\subsection{Fitting optical spectra}\label{sec:specfit}

The DR7Q and DR12Q catalogs report measurements of the quasar broad emission lines;\footnote{Emission-line measurements were performed by \citet{Shen2011} for the DR7 quasars.} however, there are no spectral measurements reported for the DR14Q catalog. Furthermore, the spectral measurements for the DR7 and DR12 quasars were performed using different algorithms. To obtain consistently measured optical/UV spectral parameters, we downloaded and fit the spectra of each of the quasars in our Sensitive sample. We employed the PyQSOFit\footnote{\url{https://github.com/legolason/PyQSOFit}} \citep{PyQSOFit} software, which is based on the code used in \citet{Shen2011}. PyQSOFit is a flexible algorithm designed to fit typical quasars (e.g.\ quasars with no broad absorption features), and it contains numerous optional features that can be used depending on the spectrum being fit. In our analysis, we always employed the {\tt{deredden}} parameter, which corrects the spectrum for Galactic extinction using the dust maps in \citet{Schlegel1998} and the extinction curve of \citet{Cardelli1989}, as well as the {\tt{and\_or\_mask}} parameter, which masked bad pixels from the pixel masks provided with the SDSS spectrum. We also adopted the redshifts published in the SDSS quasar catalogs as the systemic redshift, with the improved redshift measurements from \citet{HW2010} adopted as the systemic redshift for the DR7Q quasars.

In our analysis, we followed the methods outlined in \citet{Shen2011} to fit our quasar spectra, since both analyses systematically fit a large number of quasars, and the fitting algorithms are similar. The global continuum and the emission lines are fit separately in order to obtain a more accurate estimate for the local continuum surrounding each emission line. The global continuum was modeled using a combination of a simple power-law, a low-order polynomial, and the UV \ion{Fe}{ii} templates from \citet{Vestergaard2001}. Reliable line-free regions (at rest-frame wavelengths) are built into PyQSOFit, and we visually confirmed that these regions do not contain any apparent emission features. From the continuum fit, we computed the signal-to-noise ratio (SNR) of the line-free regions, which is used to remove low-quality spectra in our analyses.

Next we measure the properties of the \civ emission-line, which is present in all of the spectra in our redshift range. Akin to \citet{Shen2011}, we fit a local power-law continuum to the line-free region in the rest-frame window\footnote{Note that the line-free regions built into PyQSOFit do not overlap with the \civ emission line, so the continuum is not being fit in the emission-line region.} [1445,1705] \angstrom. The fitted continuum was subtracted from the \civ line region, and three broad Gaussian profiles were fit to the emission line within the window [1500,1600]~\angstrom. Both broad and narrow absorption around the emission line can have a large effect on the fitted profile; to mitigate this problem, we masked the 3$\sigma$ outliers from the spectrum smoothed by a 20-pixel boxcar filter. In cases where the absorption takes place within the emission-line, the initial fit was subtracted from the spectrum, the 3$\sigma$ outliers were masked, and the line was re-fit \citep{Shen2011}. PyQSOFit returned the EW (estimated from the multi-Gaussian model), full-width at half-maximum, line peak wavelength, and line dispersion measurements for the \civ line. We also computed the median SNR per SDSS spectral pixel surrounding the emission-line to assess spectral line quality. In this investigation, we are particularly interested in the EW and line peak wavelength measurements to compare with our X-ray data.

A subset of the quasars in our Sensitive sample have a sufficiently low redshift that the \mgii emission line is present before it exits the SDSS spectrograph range at $z\approx2.25$. For such objects, we again fitted a local continuum, using a power-law model in the wavelength range [2200,3090] \angstrom, and subtracted the UV \ion{Fe}{ii} line template of \citet{Vestergaard2001}. The continuum fit was subtracted from the emission-line region ([2700,2900] \angstrom), and the emission line fitted with three broad Gaussian profiles. In this analysis, we consider the \mgii line to be one broad line instead of a doublet because the large line-broadening of the \mgii emission line makes it difficult to measure the doublet reliably. Again, during the fitting, the 3$\sigma$ outliers are masked in the 20-pixel boxcar smoothed spectrum to reduce the impact that absorption near the \mgii emission line has on the emission-line fit.

Errors on the emission-line parameters were estimated using a Monte Carlo approach (e.g.\ \citealt{Shen2011}). A mock spectrum was created by modulating the input raw spectrum by randomly drawn values from the flux-density errors (assuming a normal distribution). The new spectrum was fit using the same techniques as above. This process was repeated 50 times, and the standard deviation of the 50 line-parameter measurements was adopted as the measurement uncertainty. We also visually inspected all of the fits to the continuum and emission lines to ensure the fits were appropriate.\footnote{From visual inspection we found that the \mgii line was best fit for quasars with $z\leq2.15$, so we restrict our analysis to this redshift range when investigating \mgii parameters.}

Furthermore, we identified BALs and mini-BALs near the \civ line that were not originally flagged in the catalogs. To identify BALs, we located the pixels that were 3$\sigma$ outliers from the ratio between the raw spectrum, convolved with a 10-pixel boxcar, and the fitted model from PyQSOFit. Next, we took the derivative of the spectrum blue-ward and red-ward of the outlier locations, and identified the pixel locations where the derivative changed sign, indicating that the absorption line had intersected the continuum. We computed the distance between the inflection points and flagged objects with $\geq30$ contiguous pixels ($\approx$2000 $\rm{km\ s^{-1}}$ for SDSS) as BAL quasars, and objects with 15--29 contiguous pixels as mini-BAL quasars. We visually inspected each spectrum to confirm the presence of the BAL or mini-BAL. We removed BAL quasars from our sample (41 quasars) since they tend to be \mbox{X-ray} weaker than typical quasars due to additional X-ray absorption in the central region (e.g.\ \citealt{Gallagher2002, Gibson2009}); however such absorption appears not to hold in mini-BAL quasars generally (e.g.\ \citealt{Wu2010, Hamann2013}), and therefore they are not removed from our sample (only $\approx$ 9\% of the sample).

Quasars with the lowest spectral quality which are not well fitted by PyQSOFit were removed from our analysis sample by imposing a constraint on the spectral SNR. We require that both the SNR of the global continuum and the median SNR per pixel of the flux in the \civ or \mgii line to be $\geq 3$. This value was chosen based on the analyses of \citet{Shen2011} and \citet{Shen2016}, which tested the effects that SNR had on the measurement of the EW and blueshift. They found little change in their measurements when the spectral quality was downsampled to a SNR $\approx3$. We report the number of quasars, X-ray detection fraction, and some of the average properties of the non-BAL quasars with good \civ and \mgii measurements in \hbox{Table \ref{tab:Samples}}. Our Full data set is described in more detail in Appendix \ref{sec:appendixB} (see Table \ref{tab:FullCatalog}), and is available online in machine-readable format.

\begin{table}
\centering
\caption{Serendipitous quasar sample properties}
\label{tab:Samples}

\begin{tabular}{l r l l l r}
\hline
Sample & $N_{\rm total}$ & $f_{\rm detected} $ &  $ \langle z \rangle $ & $ \langle \Gamma_{\rm eff} \rangle$ &  $ \langle \alpha_{\rm ox} \rangle $ \\
\thead{(1)} & \thead{(2)} & \thead{(3)} & \thead{(4)} & \thead{(5)} & \thead{(6)}  \\
\hline
Full & 2106 & 0.782 &   2.17 & 1.69 &  $-$1.46 \\
Sensitive & 753 & 0.963 &   2.16 & 1.76 & $-$1.49 \\
\ion{C}{iv}  & 637 & 0.969 &   2.17 & 1.77  & $-$1.50 \\
\ion{Mg}{ii}  & 237 & 0.992 &  1.92 & 1.80 & $-$1.50 \\
\hline
\end{tabular}

\begin{flushleft}
\footnotesize{{\it Notes:} Basic properties of the different samples used throughout this paper. Each row provides the information for the four samples considered in this work, described in Sections \ref{sec:sample_selection} and \ref{sec:Xopt_data}. Columns 2 and 3 provide number of quasars and X-ray detection fraction in each sample, and columns 4--6 report the average values of the redshift, X-ray power-law photon index, and optical-to-X-ray spectral slope of each sample.}
\end{flushleft}

\end{table}


\section{Optical \& X-ray properties}

\subsection{Optical-to-X-ray spectral slope}\label{sec:aox_defn}

\begin{figure*}
	\includegraphics[width=\textwidth]{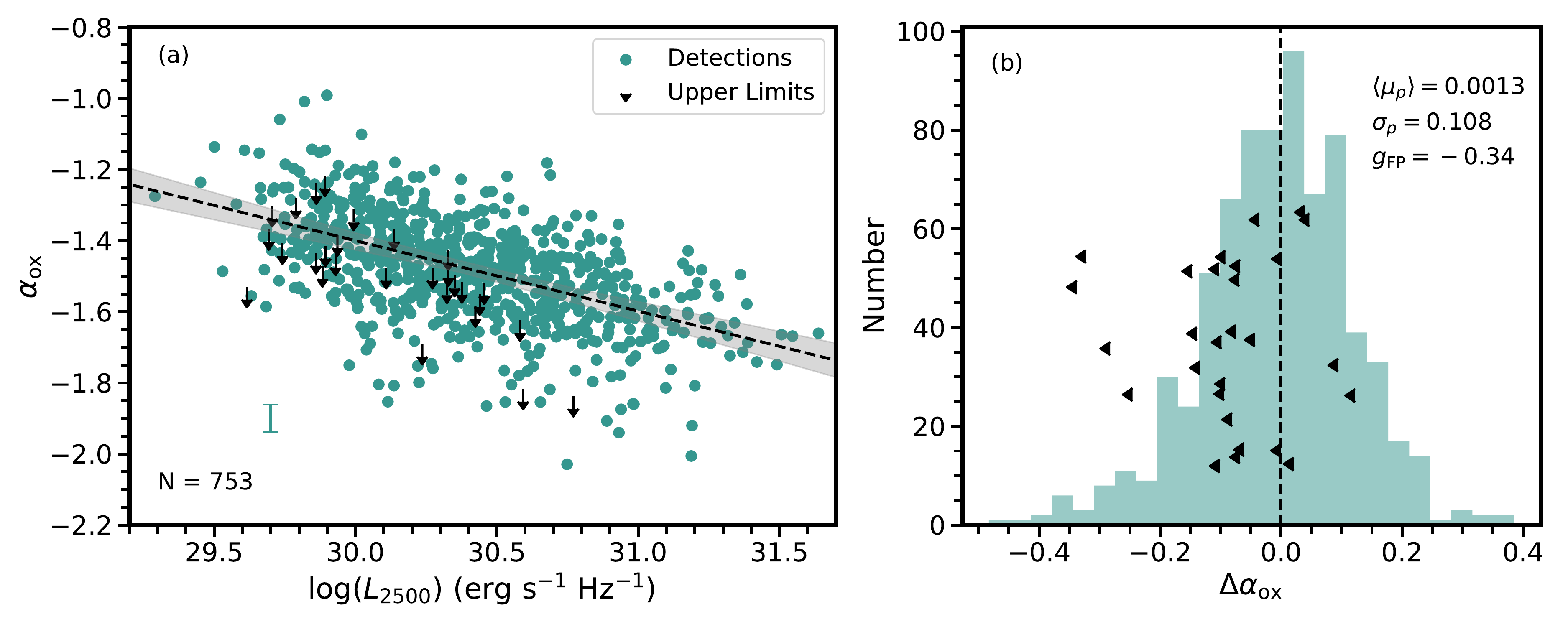}
    \caption{(a) \aox as a function of the 2500 \mbox{\AA} \ monochromatic luminosity, $L_{2500}$, for our Sensitive sample of 753 quasars where the objects flagged as BALs have been removed. Green points depict objects that are well detected above the local background, and their median errors on \aox are shown with the error bar in the lower left-hand corner.  Black downward pointing arrows represent the 90\% confidence upper limit of \aox for objects that are not detected. The black dashed line depicts the best-fit relationship between \aox and log($L_{2500}$) for our sample found from the Bayesian fitting algorithm of \citet{Kelly2007} and used to compute \daox; the 3$\sigma$ confidence interval is shown as the grey shaded region. (b) The distribution of \daox for the detected objects in this sample (green histogram) as well as the upper limits (black triangles). The vertical dashed line depicts the expected \daox given the luminosity of the quasar and the fit in panel (a). We report the mean ($\langle \mu_p \rangle$) and dispersion ($\sigma_p$) of the parent population using the likelihood method in \citet{Maccacaro1988}. The Fisher-Pearson coefficient of skewness, $g_{\rm FP}$, and the triples test demonstrates that the distribution of \daox is slightly left-hand skewed. In this investigation, objects with \mbox{\daox $< -0.2$} are considered to be X-ray weak; this relationship is used to compute \daox consistently for the WLQs.}
    \label{fig:aox_logL2500}
\end{figure*}

We computed the optical-to-X-ray spectral slope, \aox, for each of the quasars in the Full sample using the rest-frame flux densities, $f_{\rm 2keV}$ and $f_{2500}$.\footnote{Uncertainties in \aox were determined using the uncertainties in $f_{2\rm{keV}}$, which were computed by propagating the 1$\sigma$ errors on the net counts through the analysis in Section \ref{sec:Xdata_redux}. The uncertainty in $f_{2\rm{keV}}$ dominates over the uncertainty in \fopt.} In this analysis we adopt $f_{2500}$ computed using the photometric magnitudes instead of measuring it directly from the spectra because of the relatively high uncertainty in the SDSS spectral flux calibration, particularly at the red end of the spectrum \citep{Margala2016}, for DR12Q and DR14Q. While \citet{Margala2016} improved the calibration for most of the quasar spectra that comprise DR12Q, the spectra for the quasars presented in DR14Q still suffer from this large uncertainty. We opt to estimate $f_{2500}$ consistently for our Full sample rather than correct a subset of the spectra, which could introduce biases. Furthermore, for objects not detected in the \Chandra soft band, we consider $f_{\rm 2keV}$ to be an upper limit on the rest-frame flux density, even if there is a detection in the hard band (this situation is rare; see Footnote \ref{foot:limits}). 

We depict \aox as a function of the logarithm of the \hbox{2500 \angstrom} monochromatic luminosity, ${\rm{log}}_{10}(L_{2500})$, for the objects in the Sensitive sample (with BAL quasars removed) in panel (a) of \mbox{Figure \ref{fig:aox_logL2500}} (see Section 4.2 of \citealt{Brandt2015} for a review of previous studies of this relation). The best-fit relationship between these parameters was computed using the Bayesian fitting method developed in \citet{Kelly2007} and implemented in the linmix\footnote{\url{https://linmix.readthedocs.io/en/latest/maths.html}} python package. The linmix implementation fits a linear model to a univariate distribution, including errors in both dimensions, and incorporates measurements that are upper (or lower) limits (i.e.\ non-detections). The output of this algorithm returns the model, a confidence interval (CI) on the model, and an estimate of the intrinsic scatter present in the data. The best-fit relationship to our Sensitive sample of quasars\footnote{Fitting both the X-ray detected quasars as well as the 90\% confidence upper limits for the X-ray non-detections.} follows the functional form: 
\begin{equation}
\begin{split}
\alpha_{\rm{ox}} = (-0.199 \pm 0.011){\rm{log}}_{10}(L_{2500}) + (4.573\pm 0.333), \\ \sigma_{\epsilon} = 0.11, 
\end{split}
 \label{eq:aox_L2500}
\end{equation}
where $\sigma_{\epsilon}$ is the standard deviation of the intrinsic scatter.\footnote{We report the 1$\sigma$ errors on the fitted slope and intercept throughout this work.} This relationship is slightly steeper than that reported in \citet{Just2007}, who found $\alpha_{\rm{ox}} = (-0.14 \pm 0.007){\rm{log}}_{10}(L_{2500}) + (2.705 \pm 0.212)$; however, it is consistent given the intrinsic scatter and errors on our fitted parameters. Figure \ref{fig:aox_logL2500} panel (a) displays our best-fit relationship as a black dashed line; the CI on the fit is the grey shaded region. We examined the effect of the upper limits on the best-fit relationship in Equation \ref{eq:aox_L2500} by removing them from our data and refitting the trend. The new best-fit relationship, $\alpha_{\rm{ox}} = (-0.207 \pm 0.0106){\rm{log}}_{10}(L_{2500}) + (4.849\pm 0.325)$, is consistent with Equation \ref{eq:aox_L2500} within the errors, which suggests that the upper limits have little effect on our fitted trend.\footnote{We also tested the effects of using the $1\sigma$ and $3\sigma$ confidence upper limits in the fit, and similarly find relationships consistent with Equation \ref{eq:aox_L2500}.} We use Equation~\ref{eq:aox_L2500} throughout the remainder of this analysis.

The best-fit relationship in Equation \ref{eq:aox_L2500} was used to compute the X-ray weakness parameter, \daox, by taking the difference between the measured \aox and the expected \aox, which is calculated using Equation \ref{eq:aox_L2500} and the measured 2500 \angstrom luminosity of the quasar. The \daox distribution for the Sensitive sample is presented in panel (b) of Figure \ref{fig:aox_logL2500}. The parent distribution has been deconvolved from the distribution of measurement errors using the likelihood method of \citet{Maccacaro1988}. The mean and dispersion of the parent population of the \daox distribution are $\langle\mu_p\rangle = 0.0013 \pm 0.0032$ and $\sigma_p = 0.108 \pm 0.0023$, respectively. The standard deviation of \daox for the Sensitive sample ($\sigma_{\rm Sensitive}$ = 0.124) is only somewhat larger than the intrinsic dispersion computed for the parent distribution, which indicates that the spread in \daox is the dominant source of scatter over the measurement error; this result is consistent with that of the Sample B quasars from \citet{Gibson2008}. 

We tested whether the \daox distribution is symmetric about the mean by computing the Fisher-Pearson third-moment coefficient of skewness, $g_{\rm FP}$ \citep{Zwillinger2000}. The resulting value, $g_{\rm FP} = -0.34$, indicates that the distribution is mildly skewed toward negative \daox values. The triples test of symmetry \mbox{\citep{Randles1980}}, which makes no assumption about the nature of the underlying distribution, confirms that the distribution is skewed \mbox{($P_{\rm null}$ = 0.025)}. We then tested the consistency of the \daox distribution with a Gaussian distribution, finding a $p$-value for this test, $p=0.0024$, which indicates that the \daox values are not Gaussian distributed. Further investigation into the physical properties of the X-ray strong (\daox$\geq 0.2$) and X-ray weak (\daox$\leq -0.2$) tails of this distribution would be intriguing, but is beyond the scope of this work. 

Finally, we find that Equation \ref{eq:aox_L2500} best describes the relationship between the X-ray and UV emission for the data considered in this work (e.g.\ compared to the relationships of \citealt{Just2007} or \citealt{Lusso2016}), and thus provides a robust measure of \daox for the following analysis. Using the relationship in Equation \ref{eq:aox_L2500}, we estimated \aox and \daox for the Full sample. For the remainder of this work, we will investigate the relationship between the \aox (and \daox) values and the measured broad-line features such as \civew, \civ blueshift, and \mgii EW. 


\subsection{\ion{C}{\uppercase{iv}} EW relationship with \aox and \daox}\label{sec:daox_civew}

The high-ionization \civ emission line has been the subject of extensive investigation (e.g.\ \citealt{Baldwin1977, Gaskell1982,Wilkes1984,Murray1995, Leighly2004, Richards2011, Denney2012, Waters2016}), sometimes having been associated with quasar outflows. In the disk-wind model, the \civ emission-line gas directly interacts with the ionizing photons from the central engine (as opposed to the lower-ionization lines which are exposed to a reprocessed SED; e.g.\ \citealt{Leighly2004}); therefore, we expect a relation to exist between the shape of the ionizing SED (measured by \aox) and the abundance of \civ, as measured by the \civew. Furthermore, we also investigate the \daox--\civew relationship to remove the effect of the luminosity dependence of \aox. \citet{Gibson2008} did report such a correlation between \daox and \civew using their full sample (as opposed to their unbiased Sample B) of quasars (see also \citealt{Green1998}). Below, we perform a similar analysis using our unbiased sample of quasars to test the previous results and to measure better the properties of these relationships.

\begin{figure}
	\includegraphics[width=\columnwidth]{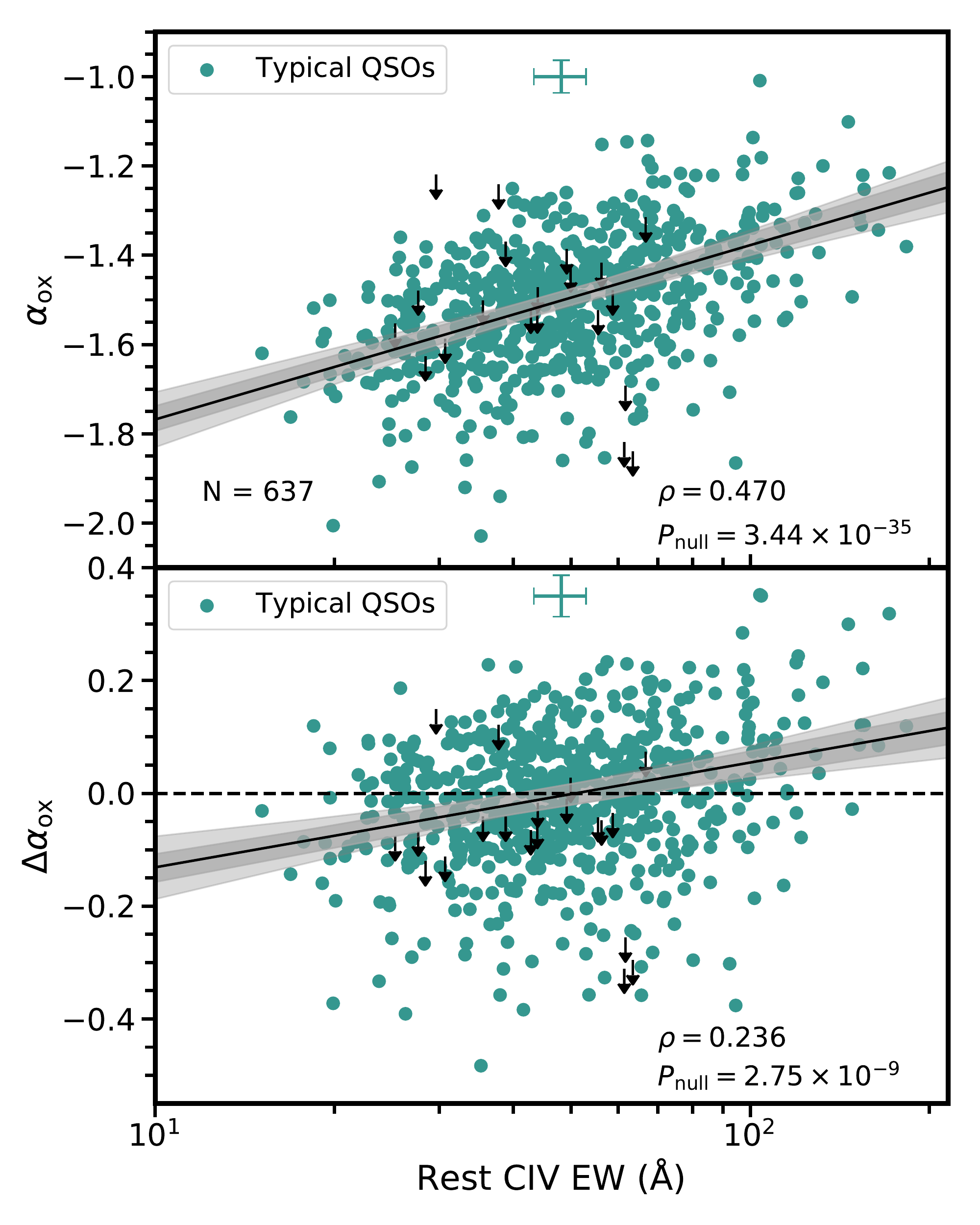}
    \caption{\aox (top panel) and \daox (bottom panel) as a function of rest-frame \civew for the 637 typical quasars (green points represent quasars with X-ray detections and black arrows depict quasars with 90\% confidence upper limits) in our \civ subsample. The best-fit relationship in each panel (black line) was found using the Bayesian fitting method of \citet{Kelly2007}, and the $1\sigma$ (dark grey) and $3\sigma$ (grey) confidence intervals are shown. Median errors on the measurements of both \aox and \civew (top panel) and \daox and \civew (bottom panel) are shown for our sample of quasars (green error bars at the top of each panel). The number of quasars in the subsample (N) is reported at the bottom left of the top panel, and the Spearman rank-order test statistic ($\rho$) and probability ($P_{\rm null}$) are reported in the bottom right of both panels. The horizontal dashed line in the bottom panel indicates the location of the expected X-ray emission given the luminosity of the quasars and the trend in Equation~\ref{eq:aox_L2500}. A more negative \daox value indicates that the quasar is X-ray weaker than expected. Here we do not adjust for the dependence of \civew on luminosity to remain consistent with previous investigations; adjusting \civew for luminosity does not change this particular result substantively (see Section \ref{sec:Baldwin}). A significant relationship exists between \aox and \civew, as well as \daox and \civew, with tight confidence intervals.}
    \label{fig:daox_civew}
\end{figure}

We investigated both the \aox--\civew and the \daox--\civew relationships using our unbiased \civ subsample of quasars, depicted in the top and bottom panels of Figure \ref{fig:daox_civew}, respectively. To test for correlations in these two spaces, we performed a Spearman rank-order test, as implemented in the Astronomy Survival Analysis package (ASURV; \citealt{Feigelson1985, ASURV}). This test reports significant correlations (at the $\geq99.99\%$ level) between \aox and \civew as well as \daox and \civew. The width of the confidence interval on the Spearman rank-order correlation coefficient, $\rho$, is inversely related to the number of objects in the sample. Since the \civ subsample contains a large number of quasars (637), the correlation coefficient can be determined with high confidence. To quantify the effect that the measurement errors of the data have on the correlation coefficient, we employ a Monte Carlo method (e.g.\ \citealt{Curran2015}). We resampled the data 5000 times by adding random offsets,\footnote{The offset values were randomly drawn from a normal distribution with a mean of zero and a standard deviation equal to the measurement error for each data point.} recomputed the Spearman test, and use the standard deviation, $\sigma_{\rho}$, of the new coefficients as limits on the original correlation coefficient (both $\rho$ and $\sigma_{\rho}$ are reported in Table \ref{tab:twoSamplestats}). We find that the errors on the data points have little effect on the correlation coefficient for both the \aox--\civew and \daox--\civew parameter spaces; therefore, the correlations are robust and a relationship can be meaningfully fit to these data. Correlations investigated later in this paper will similarly be assessed for robustness using this Monte Carlo approach (see column 6 of Table \ref{tab:twoSamplestats}). Using the prescription of \mbox{\citet{Kelly2007}} the best-fit linear relationships are
\begin{equation} 
\begin{split}
\alpha_{\rm{ox}} = (0.390 \pm 0.027)\rm{log_{10}(\civ \ EW)} - (2.158 \pm 0.047) \\ \sigma_{\epsilon} =  0.11, \\
\end{split}
 \label{eq:aox_civew} 
\end{equation}

\begin{equation} 
\begin{split}
\Delta\alpha_{\rm{ox}} = (0.185 \pm 0.025)\rm{log_{10}(\civ \ EW)} - (0.315 \pm 0.043) \\ \sigma_{\epsilon} =  0.10, 
\end{split}
 \label{eq:daox_civew} 
\end{equation}
where both \aox and \daox are fitted with respect to $\rm log_{10}$(\civew), and $\sigma_{\epsilon}$ is again the standard deviation of the intrinsic scatter. The best-fit relations in Equations \ref{eq:aox_civew} and \ref{eq:daox_civew}, along with the 1$\sigma$ and 3$\sigma$ CI, were fitted using quasars with \mbox{\civew $\geq 15$ \angstrom} so that the WLQs were excluded from this analysis.\footnote{The WLQs span a different range in luminosity and likely have different \daox properties than the typical quasars, so including them might bias these relationships. Furthermore, we wish to investigate if any trends from the typical quasars can predict the X-ray weakness of the WLQs, so including the WLQs in the fit is inappropriate.} Equation \ref{eq:aox_civew} reports a clear positive correlation between the ionizing SED of the quasar and the \civew (discussed in further detail in Section \ref{sec:qso_corr}). Equation \ref{eq:daox_civew} demonstrates that the luminosity dependence of \aox does not artificially generate a relationship with \civew. While the slope of the linear fit in Equation \ref{eq:daox_civew} is slightly less steep than that reported in \citet{Gibson2008}, the best-fit relationships are consistent within the CI. We have improved the confidence in these relationships primarily because our \civ subsample contains a larger number of objects with a larger fraction detected in X-rays (637 quasars; $\approx$~96\% X-ray detected) compared to the full sample in \citet{Gibson2008} (433 quasars; $\approx$ 81\% X-ray detected); therefore, our fit is less reliant on X-ray upper limits. Acquiring more data at large \civew (\civew $>150$ \angstrom) could additionally constrain this fit, as it would expand the dynamic range of the data in this parameter space.

\begin{figure*}
	\includegraphics[width=0.9\textwidth]{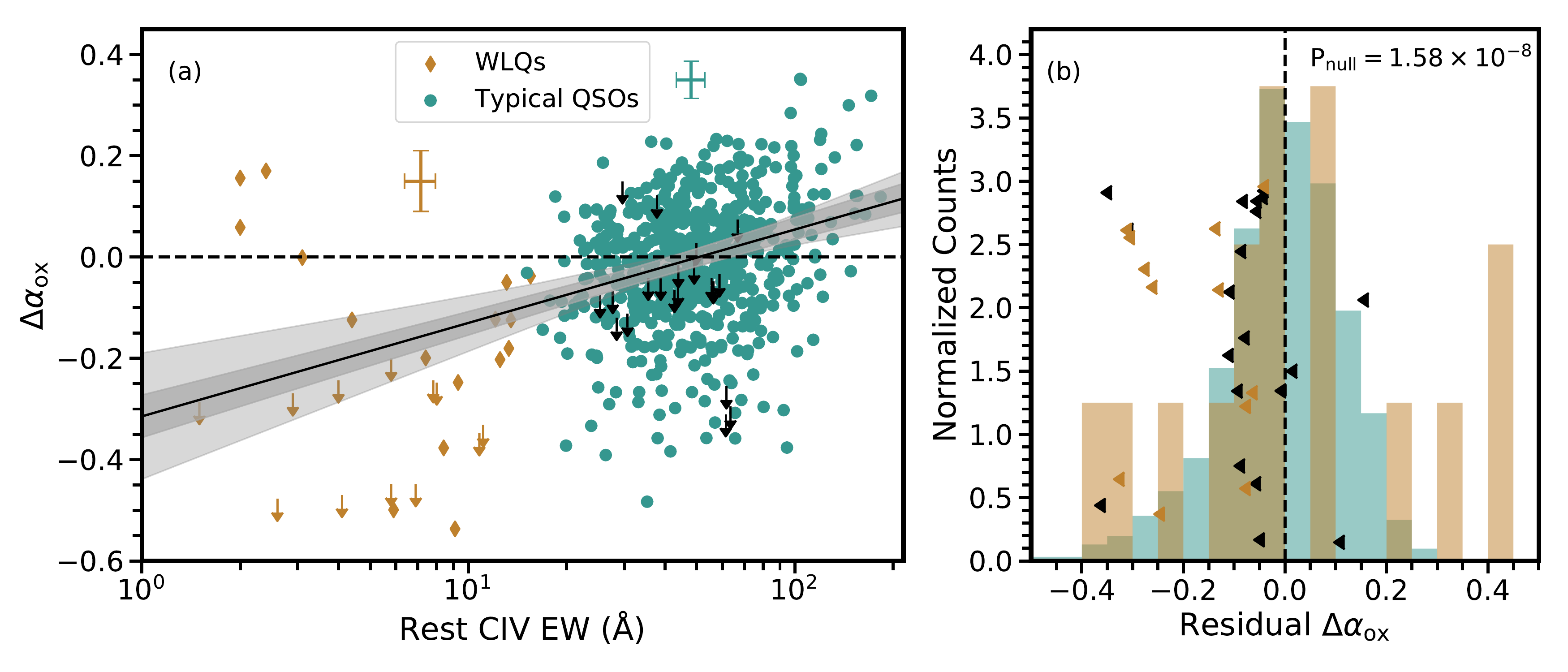}
    \caption{(a) Same as the bottom panel of Figure \ref{fig:daox_civew}, however the WLQs (brown points; brown arrows depict quasars with X-ray upper limits) from \citet{Ni2018} are included. If the WLQs behaved in the same manner as the typical quasars in this space, we would expect a comparable amount of dispersion around the best-fit trend (black line). The WLQs, however, display a much larger scatter around this line, and many are significantly X-ray weaker than predicted by the typical quasars. (b) The distribution of residual \daox around the best-fit relationship for the WLQs (brown histogram) and normal quasars (green histogram). The vertical dashed line indicates the best-fit relationship from panel (a). A Peto-Prentice two-sample test reveals that the WLQs and the typical quasars are not drawn from the same distribution in \mbox{\daox--\civew} space at the $\approx$5.8$\sigma$ level.}
    \label{fig:resid_med}
\end{figure*}

The best-fit relationship in Equation \ref{eq:daox_civew} was extrapolated to low values of \civew to assess the expected X-ray weakness of the WLQs using \daox.\footnote{We do not extrapolate Equation \ref{eq:aox_civew} to predict \aox for the WLQs because the typical quasars and WLQs span significantly different luminosity ranges (e.g.\ see Figure \ref{fig:Mi_z}).} Panel (a) of Figure \ref{fig:resid_med} compares this prediction with the distribution of the WLQs from \citet{Ni2018} in this parameter space. To compare consistently measured \daox values, we compute \daox for the WLQs using Equation \ref{eq:aox_L2500} and their reported values of \aox (from \citealt{Ni2018}, or references therein) and $L_{2500}$, which they adopt from \citet{Shen2011} (and are consistent with our values). While some of the WLQs can be predicted by the extrapolation of the typical quasar population, there is a larger apparent scatter around the relationship for the WLQ population compared to the typical quasars. To test this result, we computed the residual \daox from the best-fit relation for both the typical quasars in our \civ subset and the WLQs. Panel (b) of Figure \ref{fig:resid_med} shows the distribution of the residual \daox for both populations. We performed a Peto-Prentice test \citep{Prentice1979} using the {\tt{R}} package {\tt{EnvStats}} \citep{EnvStats} to quantify the difference, if any, between the two distributions. We adopted the Peto-Prentice test because it accounts for left-censored data. The two populations are not drawn from the same distribution ($P_{\rm null} = 1.58 \times 10^{-8}$; see Table \ref{tab:twoSamplestats}). The WLQs tend to be X-ray weaker, in general, than expected from the relationship for the typical quasars. However, there are some notably X-ray strong WLQs as well. 


\subsection{\ion{C}{\uppercase{iv}} blueshift relationship with \aox and \daox}\label{sec:daox_civbl}

Another reason that \civ line emission has been associated with quasar winds is the large blueshift that it can exhibit with respect to the systemic redshift of the quasar (e.g.\ \citealt{Richards2011,Coatman2017}). We therefore investigated whether there is a correlation between \aox (and \daox) and \civ blueshift, to further test the disk-wind model, and whether there is a connection between the typical quasars and WLQs. We note that the relevant \civ emission-line blueshift velocities are small compared to the speed of light;\footnote{The maximum measured velocity is $\approx 4000$ km s$^{-1}$ in our quasar sample, which is sufficiently small with respect to the speed of light to ignore the second-order velocity term when computing blueshifts. Most of our quasars have measured velocities $<$ 1500 km s$^{-1}$.} therefore, we define the \civ blueshift as $c(1549.06 - \lambda_{\rm peak})/1549.06$, where $\lambda_{\rm peak}$ is the measured peak of the emission line (in \angstrom), 1549.06 \angstrom is the laboratory wavelength of the \civ emission line (see Table 4 of \citealt{Vandenberk2001}), and $c$ is the speed of light. According to this definition, quasars with {\emph{larger}} blueshifts have {\emph{positive}} ``\civ blueshift'' values.  We measure blueshifts in units of km s$^{-1}$.

\begin{figure}
	\includegraphics[width=\columnwidth]{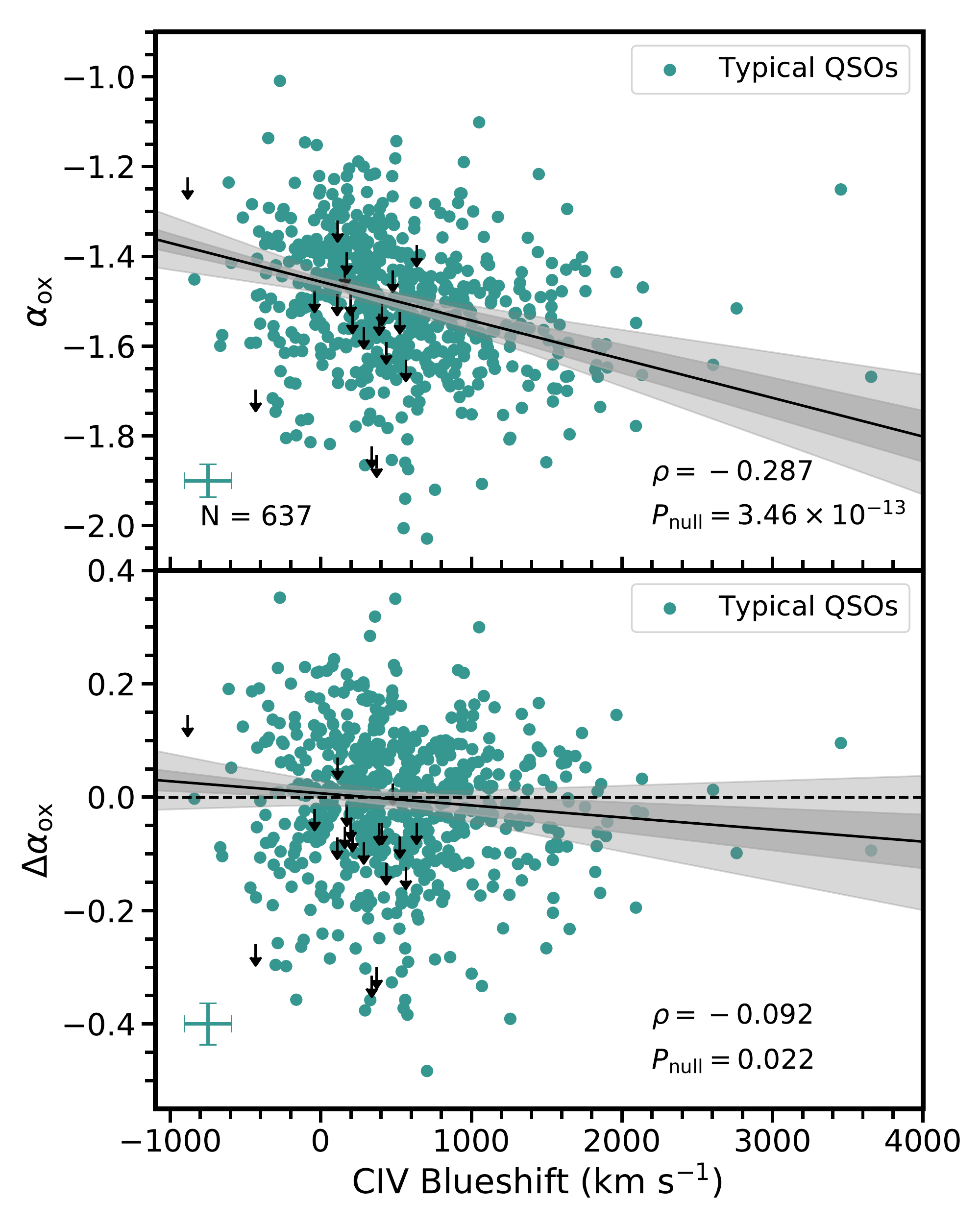}
    \caption{\aox as a function of \civ blueshift (top panel), and \daox as a function of \civ blueshift (bottom panel) for the typical quasars (green points, black arrows). Larger blueshifts are represented by positive values. The best-fit relationship (solid black line) in each panel is depicted by the grey shaded 3$\sigma$ confidence interval (the 1$\sigma$ confidence interval is depicted by the dark grey region). At the bottom left corner of each panel, we depict the mean errors in both parameters as green error bars; the horizontal dashed line in the bottom panel represents the expected \aox given the quasar luminosity and Equation \ref{eq:aox_L2500}. The number of quasars in the subsample (N) is reported at the bottom left of the top panel, and the Spearman rank-order test statistic ($\rho$) and probability of no correlation ($P_{\rm null}$) are reported in the lower right of each panel. We find a strong correlation between \aox and \civ blueshift, whereas there is only a slight correlation between \daox and \civ blueshift. As was seen in Figure \ref{fig:daox_civew}, the luminosity dependence of the \civ blueshift does not change these results (see Section \ref{sec:Baldwin}).}
    \label{fig:aoxdaox_blueshift}
\end{figure}

The top panel of Figure \ref{fig:aoxdaox_blueshift} depicts \aox as a function of \civ blueshift.\footnote{We also investigated \aox vs. log$_{10}$(\civ blueshift + 2000 km s$^{-1}$), where the addition of 2000 km s$^{-1}$ ensures that the argument of the logarithm is not negative, and our conclusions remain the same.} We again perform a Spearman rank-order test to determine if these parameters are correlated, and find a significant correlation ($P_{\rm null} = 3.46\times10^{-13}$; see Table \ref{tab:twoSamplestats}). To remove the luminosity dependence of \aox, we also investigate \daox as a function of \civ blueshift as depicted in bottom panel of Figure \ref{fig:aoxdaox_blueshift}. We find a weak correlation between these two parameters (see Table~\ref{tab:twoSamplestats}).\footnote{We find a stronger correlation ($\rho = -0.16$, $P_{\rm null} = 2.5\times10^{-5}$) between \daox and \civ blueshift than is reported in Table \ref{tab:twoSamplestats} when using \daox values calculated using the \aox--$L_{2500}$ relation from \citet{Lusso2016}. This suggests that the existence of the \daox and \civ blueshift correlation is independent of the model used to compute \daox.} Again, we fit a linear relationship to our typical quasars, and find that the best-fit relationships are
\begin{equation} 
\begin{split}
\alpha_{\rm{ox}} = (-8.656 \pm 1.256)\times 10^{-5}(\civ\ \rm{blueshift}) \\ + (1.456 \pm 0.008), \sigma_{\epsilon} =  0.13,
\end{split}
 \label{eq:aox_civbl}
\end{equation}

\begin{equation} 
\begin{split}
\Delta\alpha_{\rm{ox}} = (-2.10 \pm 1.07)\times 10^{-5}(\civ\ \rm{blueshift}) \\ + (0.007 \pm 0.007), \sigma_{\epsilon} =  0.11, 
\end{split}
 \label{eq:daox_civbl}
\end{equation}
As depicted in Figure \ref{fig:aoxdaox_blueshift}, \aox and \daox are both negatively correlated with \civ blueshift, which implies that softer SEDs generate higher velocity outflows; however, the fitted \daox--\civ blueshift relationship is significantly flatter than the \aox--\civ blueshift relationship. As an additional test of kinematics, we investigated \aox (and \daox) as a function of the full-width at half-maximum (FWHM) of the \civ emission line, but found no significant correlation in this space.

\begin{figure}
	\includegraphics[width=\columnwidth]{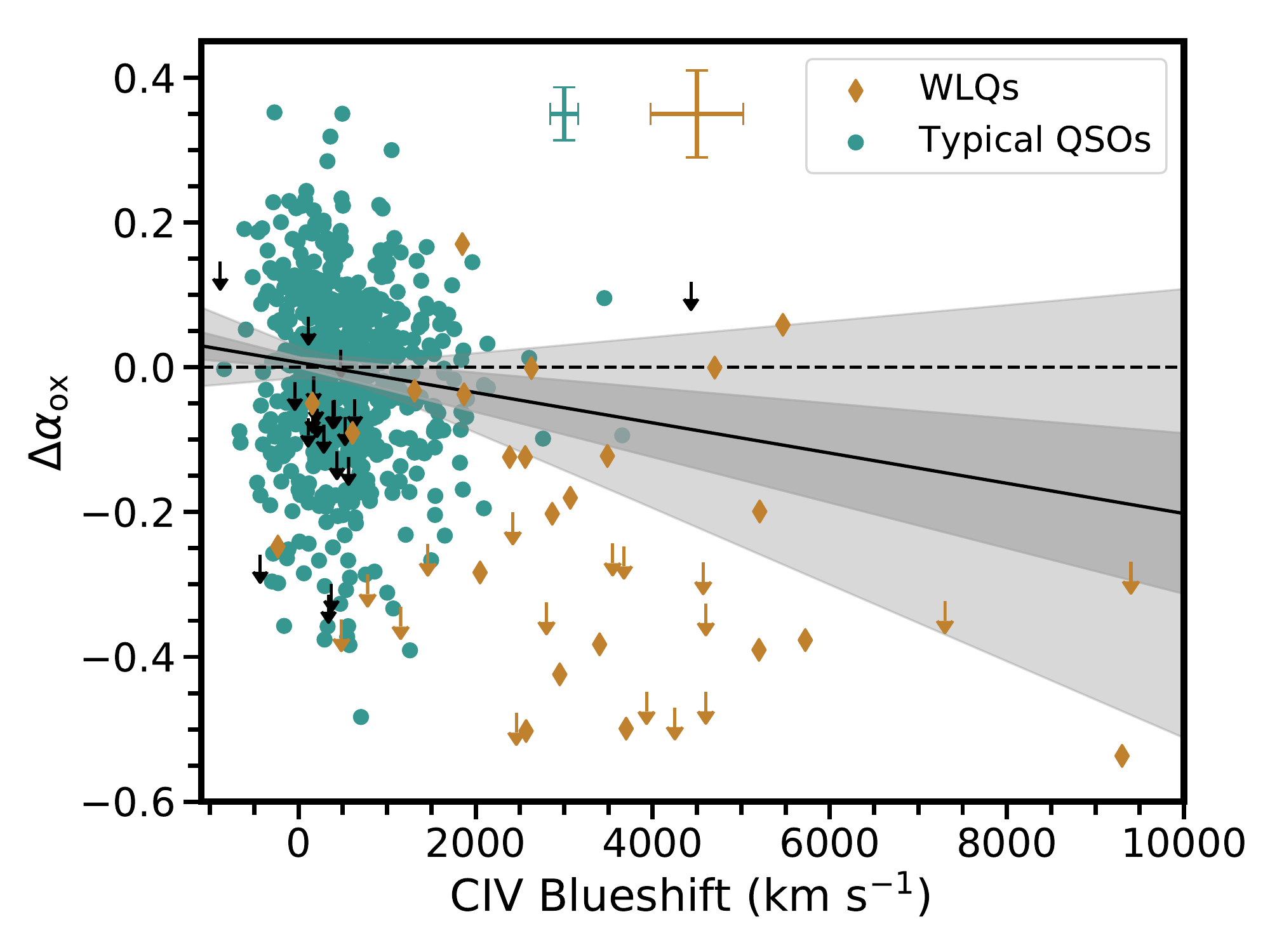}
    \caption{Similar to the bottom panel of Figure \ref{fig:aoxdaox_blueshift}, however we include the \daox and \civ blueshift values for the WLQs (brown diamonds and arrows). Error bars are depicted near the top of the plot. The WLQs clearly tend to display much larger blueshifts than those of the typical quasars. We extrapolate the best-fit relationship from Equation \ref{eq:daox_civbl} to try to predict the \mbox{X-ray} weakness of the WLQs; however, the WLQs tend to be X-ray weaker than the \daox relationship with \civ blueshift of the typical quasars predicts.}
    \label{fig:daox_blueshift}
\end{figure}

We additionally depict the \civ blueshift values of the WLQs in \mbox{Figure \ref{fig:daox_blueshift}} which tend to be much larger than those of the rest of the quasar population (see also \citealt{Ni2018} and references therein). Also apparent in Figure \ref{fig:daox_blueshift} is that the $3\sigma$ confidence interval on Equation \ref{eq:daox_civbl} greatly expands which we attribute to the large scatter in \daox.\footnote{For reference, we also depict the $1\sigma$ CI which is not nearly as large.} Despite the large $3\sigma$ CI (which is consistent with no correlation in this space), we extrapolated the fit from the typical quasars to larger blueshifts and computed the residual \daox for both the typical quasars and the WLQs. A Peto-Prentice test indicates that the values for the WLQs cannot be predicted by the typical quasar \daox--\civ blueshift relation (see Table \ref{tab:twoSamplestats}). Further discussion of the \daox--\civ blueshift relationship is provided in Section \ref{sec:qso_corr}.

\begin{table*}
\centering
\caption{Statistical tests for correlations and of similarity in each parameter space}
\label{tab:twoSamplestats}

\begin{tabular}{l r r c c r c c c}
\hline
 &  & & & &\multicolumn{2}{c}{Spearman rank-order} & \multicolumn{2}{l}{Peto-Prentice} \\ \cline{6-7} \cline{8-9}
\thead{Parameter space} & \thead{$N_{\rm typical}^{\rm detected}$} & \thead{$N_{\rm typical}^{\rm limit}$} & $N_{\rm WLQ}^{\rm detected}$ & $N_{\rm WLQ}^{\rm limit}$ & \thead{$\rho \ (\sigma_{\rho})^{\rm a}$} &  $P_{\rm null}^{\rho}$ ($P_{\rm null}^{\rho + \sigma_{\rho}}$, $P_{\rm null}^{\rho - \sigma_{\rho}}$)$^{\rm b}$& $\sigma^{\rm c}$ &  $P_{\rm null}$  \\ 
\thead{(1)} & \thead{(2)} & \thead{(3)} & \thead{(4)} & \thead{(5)} & \thead{(6)} & \thead{(7)} & \thead{(8)} & \thead{(9)} \\

\hline
\aox--\civew & 617 & 20 &  23 & 16 & 0.470 (0.016) & 3.44$\times 10^{-35}$ (6.85$\times 10^{-38}$, 1.05$\times 10^{-32}$) & --$^{\rm d}$ & -- \\

\daox--\civew & 617 & 20 &  23 & 16 & 0.236 (0.018) & 2.75$\times 10^{-9}$ (1.53$\times 10^{-10}$, 4.50$\times 10^{-8}$)   & 5.65 & 1.58$\times 10^{-8}$ \\

\aox--\civ blueshift & 617 & 20 &  23 & 16 & $-$0.287 (0.021) & 3.46$\times 10^{-13}$ (5.03$\times 10^{-15}$, 1.88$\times 10^{-11}$)   & -- & --  \\

\daox--\civ blueshift$^{\rm e}$ & 617 & 20 &  23 & 16 & $-$0.092 (0.023) & 0.022 (0.0042, 0.087)  & 5.98 & 2.23$\times 10^{-9}$  \\

\aox--\mgiiew  & 235 & 2 &  16 & 12 & 0.404 (0.032) & 1.19$\times 10^{-10}$ (2.53$\times 10^{-12}$, 3.99$\times 10^{-9}$)   & -- & --  \\

\daox--\mgiiew  & 235 & 2 &  16 & 12 & 0.188 (0.035) & 0.004 (5.74$\times 10^{-4}$, 0.019)  & 6.87 & 6.47$\times 10^{-12}$  \\

\aox--$\rm{R_{\civ}}$ & 235 & 2 &  16 & 12 & 0.212 (0.039) & 0.001 (1.00$\times 10^{-4}$, 0.0079)  & -- & --  \\

\daox--$\rm{R_{\civ}}$ & 235 & 2 &  16 & 12 & 0.184 (0.040) & 0.004 (5.41$\times 10^{-4}$, 0.027)  & 4.00 & 6.09$\times 10^{-5}$  \\
\hline
\end{tabular}

\begin{flushleft}
\footnotesize{{\it Notes:} We report the number of typical quasars and WLQs (detected and undetected) used to fit the relationships in each parameter space (see columns 1--5). In the next four columns, we show the test statistic and $P_{\rm null}$ for the Spearman rank-order correlation test and the Peto-Prentice two-sample similarity test (only for the \daox relationships). Correlation tests were performed with the typical quasars to assess if a relationship should be fitted to the data. The Peto-Prentice test reports the similarity of the residual \daox between the WLQs and the typical quasars given the best-fit relationship. The WLQs tend to be X-ray weaker than predicted by all of the quasar relationships we investigated.  \\ 
$^{\rm a}$ Test statistic, $\rho$, for the Spearman rank-order correlation test. The value in parentheses reports the $1\sigma$ uncertainty on the test statistic, which was calculated using a Monte Carlo method.  \\ 
$^{\rm b}$ Null hypothesis probability computed using the Spearman rank-order correlation test statistic, $\rho$. The values in the parentheses report the $P_{\rm null}$ values when the $1\sigma$ uncertainty is added to and subtracted from the test statistic, respectively. \\ 
$^{\rm c}$ Test statistic for the Peto-Prentice test \\ 
$^{\rm d}$ We do not extrapolate the \aox relationships to predict the \aox for the WLQs since the typical quasars and WLQs do not span the same luminosity range.\\
$^{\rm e}$ The result of the Peto-Prentice test returns $P_{\rm null}=0$, so we report the result of a Peto-Peto test.\\
}
\end{flushleft}

\end{table*}


\subsection{The Baldwin effect and the \texorpdfstring{\ion{C}{\uppercase{iv}} EW--\ion{C}{\uppercase{iv}}}\ blueshift relationship}\label{sec:Baldwin}

\begin{figure*}
	\includegraphics[width=0.9\textwidth]{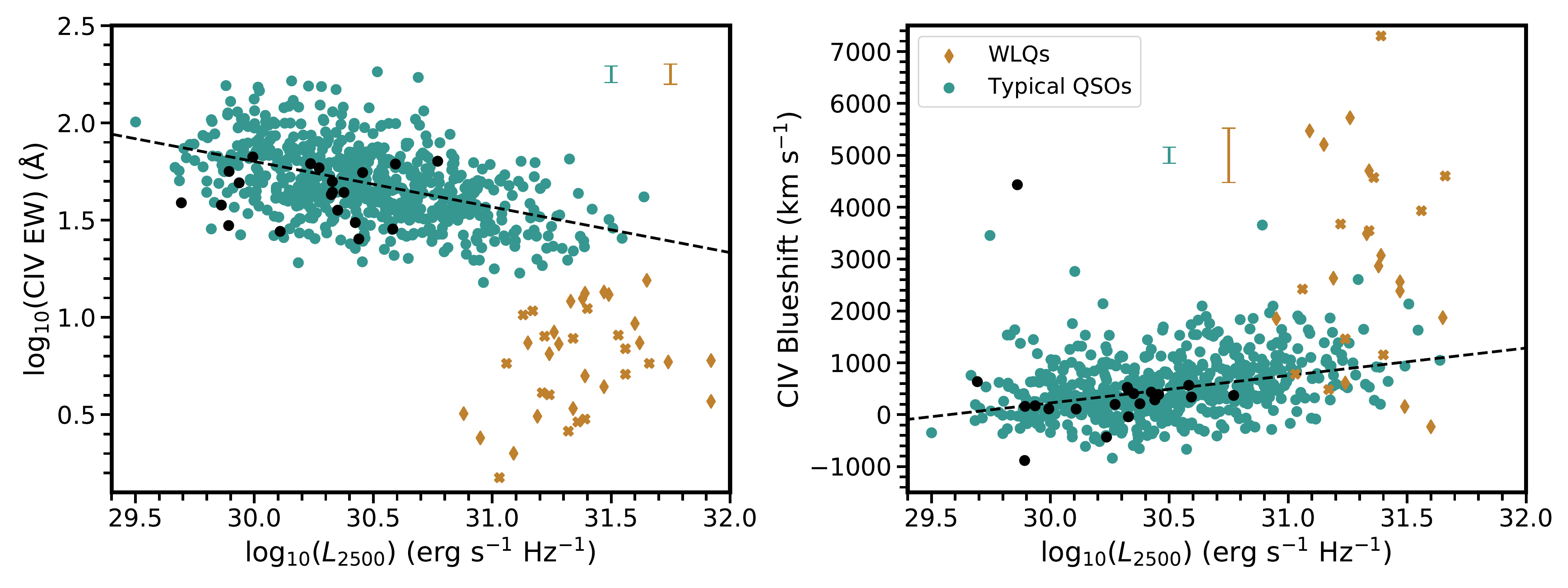}
    \caption{Dependence of log$_{10}$(\civew) (left) and \civ blueshift (right) on the 2500 \angstrom monochromatic luminosity for the typical quasars (green points; black points represent quasars with X-ray upper limits) and WLQs (brown diamonds; brown crosses indicate WLQs with X-ray upper limits). The black dashed lines depict the best-fit relationship (using the method of \citealt{Kelly2007}) between log$_{10}$(\civew) and log$_{10}(L_{2500})$ (left; see Equation \ref{eq:civew_L2500}) and \civ blueshift and log$_{10}(L_{2500})$ (right; see Equation \ref{eq:civbl_L2500}). As for the calculation of \daox, we use these best-fit relations to compute a luminosity corrected $\Delta{\rm{log}}_{10}$(\civew) and $\Delta$(\civ blueshift) to account for the difference in luminosity ranges between the WLQs and our normal quasar sample. Median errors in the \civew (left) and \civ blueshift (right) of the typical quasars (WLQs) are indicated by the green (brown) error bars. We use these luminosity-adjusted values to investigate the relationship between these two parameters for the typical quasars and the WLQs. Recall that only bright WLQs were selected for targeted X-ray observations and therefore are biased toward larger luminosity values compared to the typical quasars due to this selection effect.}
    \label{fig:civ_L2500}
\end{figure*}

In the previous sections, we have directly compared \daox and the \civ parameters of our typical quasars to those of the WLQs; however, the two sample populations are not inherently the same. Rather, both the \civew and \civ blueshift depend on the quasar luminosity, as shown in Figure \ref{fig:civ_L2500}. The observed \civew dependence on luminosity (the Baldwin effect; \citealt{Baldwin1977}) is a well-known phenomenon; however, several authors have demonstrated that it is likely a secondary effect, and that the dependence on the Eddington ratio is stronger, using samples of low-$z$ \citep{Baskin2004} and high-$z$ \citep{Shemmer2015} quasars. 

In this work, we only considered the empirical \mbox{\civew--luminosity} dependence, since estimating the Eddington ratio yields highly uncertain results. To test if the Baldwin effect has any dependence on the \daox--\civew relationship, we calculated the best-fit relationship between log$_{10}$(\civew) and 2500 \angstrom luminosity:

\begin{equation} 
\begin{split}
\rm{log_{10}(\civ \ EW)} = (-0.234 \pm 0.017){\rm log}_{10}({\it L}_{2500}) \\ + (8.824 \pm 0.529), \sigma_{\epsilon} = 0.16 \: \angstrom,
\end{split}
\label{eq:civew_L2500}
\end{equation}
shown in the left-hand panel of Figure \ref{fig:civ_L2500}. Our best-fit trend is consistent with previous results (e.g.\ \citealt{Green2001}), within the error. Next, we calculated the ``luminosity-adjusted'' \civew (using Equation \ref{eq:civew_L2500}) to account for the Baldwin effect using the same basic approach as in Section \ref{sec:aox_defn} when computing \daox from \aox, the 2500 \angstrom luminosity, and Equation \ref{eq:aox_L2500}. We repeated the analysis in Section \ref{sec:daox_civew} with the luminosity-adjusted \civew, but found no difference in the conclusions. 

Similarly, for \civ blueshift, we fit the luminosity dependence (see \citealt{Shen2016}):

\begin{equation}
\begin{split}
\rm{\civ \ Blueshift} = (529.465\pm 50.415)\rm{log}_{10}({\it L}_{2500}) \\ - (15659.650 \pm 1536.950), \sigma_{\epsilon} = 434 \:{\rm km\,s^{-1}},
\end{split}
\label{eq:civbl_L2500}
\end{equation}
and depict the relationship in the right panel of Figure \ref{fig:civ_L2500}. We compare this result to the result in \citet{Shen2016}, and find a similar fit. As before, the $\Delta$(\civ blueshift) was computed from the \civ blueshift and best-fit relationship in Equation \ref{eq:civbl_L2500}. We found no difference in the conclusions upon re-evaluating Section \ref{sec:daox_civbl} with the luminosity-adjusted blueshift values.

\begin{figure}
	\includegraphics[width=\columnwidth]{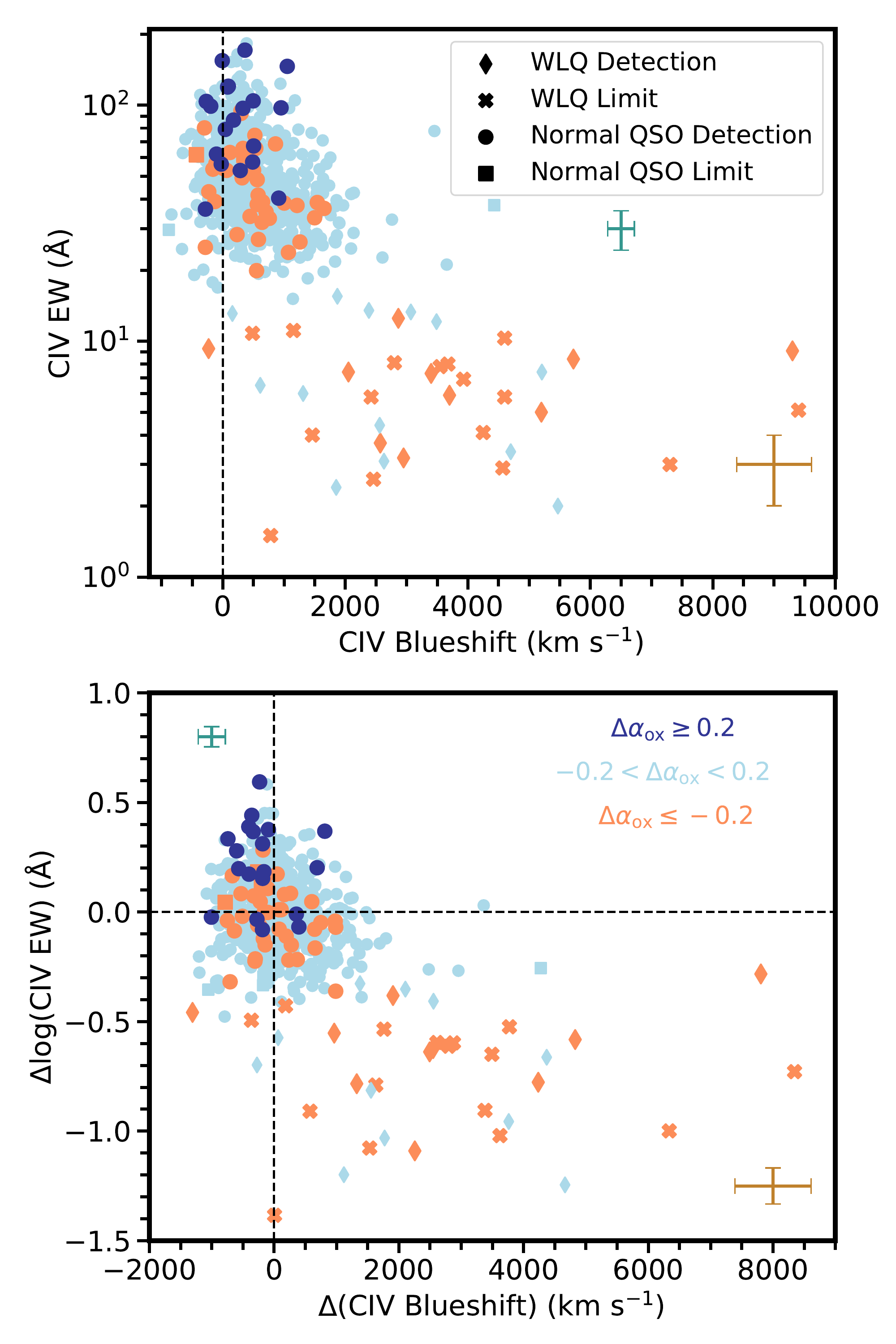}
    \caption{Top: The \civew is depicted with respect to the \civ blueshift, where X-ray detected typical quasars (WLQs) are shown as points (diamonds), and quasars with 90\% confidence X-ray upper limits as squares (crosses). Additionally, the samples are divided into X-ray weak \mbox{(\daox $\leq -0.2$; orange)}, X-ray normal ($-0.2 <$ \daox $< 0.2$; light blue), and X-ray strong (\daox $\geq 0.2$; dark blue) objects. Mean measurement errors are depicted for the typical quasars (green; middle right) and WLQs (brown; lower right), and the vertical dashed line represents a zero velocity blueshift. The X-ray strong and X-ray weak quasars tend to occupy different locations in this space; however, the distribution of X-ray normal quasars tends to envelop these X-ray outliers (see Table \ref{tab:lumadj_civ}). The X-ray weak typical quasars have blueshift values that are slightly larger than their X-ray strong counterparts; however, the X-ray weak WLQs tend to have much larger blueshifts. Bottom: Same as the top panel, however we depict the luminosity-adjusted \civew, $\Delta$log(\civew), with respect to luminosity-adjusted \civ blueshift, \mbox{$\Delta$(\civ blueshift)}. Mean error bars are located in the top-left corner (typical quasars) and bottom-right corner (WLQs), and the vertical and horizontal dashed lines represent the expected \civ blueshift and \civew according to Equations \ref{eq:civew_L2500} and \ref{eq:civbl_L2500}. We find that the X-ray subsample comparisons are similar to those in the top panel (see Table \ref{tab:lumadj_civ}).}
    \label{fig:civew_blueshift}
\end{figure}

Using our large, unbiased sample of quasars, we investigated \civew as a function of \civ blueshift. Comparing these two parameters, along with their X-ray properties, has been used to frame quasars in the context of the disk--wind model (e.g.\ see \citealt{Richards2011,Wu2011}); therefore, we can use our sample to test previous results. Since both the \civew and \civ blueshift parameters depend on the luminosity, we chose also to investigate their luminosity-adjusted values to remove any biases caused by this dependence. Figure \ref{fig:civew_blueshift} displays the \civew as a function of \civ blueshift in the top panel and \mbox{$\Delta{\rm{log}}_{10}$(\civew)} as a function of \mbox{$\Delta$(\civ blueshift)} in the bottom panel. In both panels, we split our sample into \mbox{X-ray} strong \mbox{(\daox$\geq0.2$)}, X-ray normal \mbox{($-0.2<$\daox$<0.2$)}, and X-ray weak \mbox{(\daox$\leq-0.2$)} subsets, and present the statistics of these three samples in Table \ref{tab:lumadj_civ}. The X-ray strong quasars have both smaller \civ blueshifts and larger \civew compared to the \mbox{X-ray} weak quasars (see Table \ref{tab:lumadj_civ}). Their luminosity-adjusted counterparts also follow a similar trend, albeit with less scatter (measured by the interquartile range; IQR). This result is consistent with previous studies from \citet{Wu2011}, who investigated a similar parameter space using the Sample B quasars from \mbox{\citet{Gibson2008}}.\footnote{The analysis in \citet{Wu2011} compared \civew with \civ blueshift but not their luminosity-adjusted values.} Both panels in Figure~\ref{fig:civew_blueshift} also present the WLQs from \mbox{\citet{Ni2018}}, which are split into the same \mbox{X-ray} bins.\footnote{None of the WLQs is X-ray strong; see Figure \ref{fig:resid_med}.} In addition to having smaller \civew than our typical quasars, the WLQs generally display much larger \civ blueshifts for both the X-ray weak and X-ray normal samples. 

\begin{figure}
	\includegraphics[width=\columnwidth]{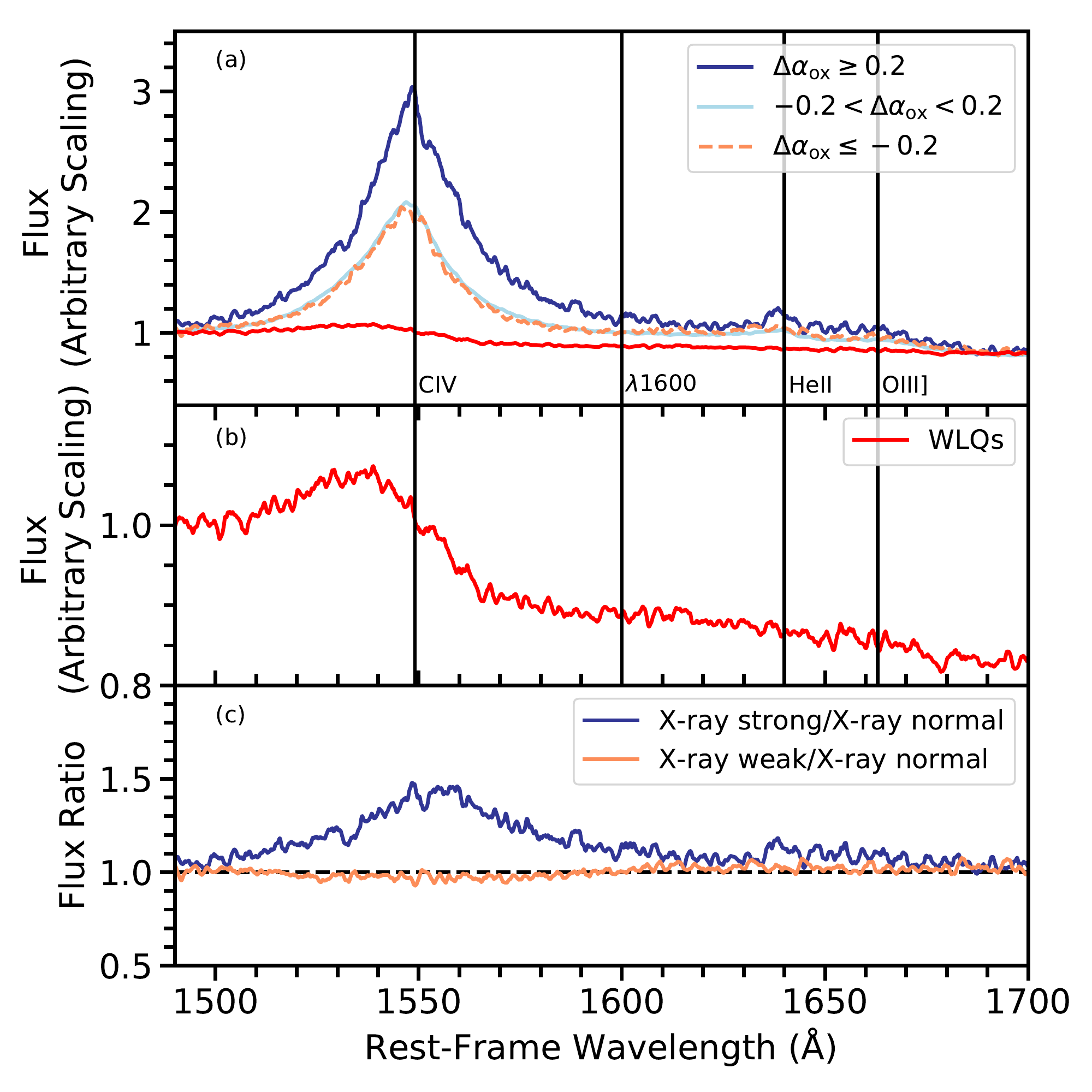}
    \caption{(a) The median stacked \civ emission-line profiles for the X-ray strong (\daox $\geq 0.2$; dark blue line), X-ray normal ($-0.2<$\daox$<0.2$; light blue line), and X-ray weak (\daox $\leq -0.2$; orange dashed line) typical quasars, along with the median stacked \civ profile for the WLQs (red line). The vertical solid lines show the rest-frame wavelengths of the \civ, $\lambda1600$, \ion{He}{ii}, and \ion{O}{iii]} emission features. Consistent with the reported values in Table \ref{tab:lumadj_civ}, we find that the X-ray strong typical quasars have a much stronger \civ profile, and exhibit smaller \civ blueshift, compared to the X-ray normal and X-ray weak typical quasars (which have similar median values; see Table~\ref{tab:lumadj_civ}). (b) The median stacked spectrum of the WLQs, shown in more detail, clearly has weaker \civ emission and has a much larger blueshift than for the typical quasars. The \civ line profile is also much more asymmetric than for the typical quasars. (c) The ratio of the X-ray strong to X-ray normal typical quasars (dark blue line) and the X-ray weak to X-ray normal typical quasars (orange line). The ratio of the X-ray weak to X-ray normal quasar spectra is nearly unity (black dashed line) and the ratio of the X-ray strong to X-ray normal quasar spectra depicts the fact that the X-ray strong subsample has the largest \civ kurtosis.}
    \label{fig:linestack}
\end{figure}

We visually compared the \civ emission-line profiles for the X-ray strong, X-ray normal, and X-ray weak typical quasars (as well as the WLQs) using their median stacked spectra (e.g.\ \citealt{Vandenberk2001}) as depicted in panel (a) of Figure \ref{fig:linestack}. To perform the median stack, the spectra in each subsample were first corrected for Galactic extinction and were cleaned using the $3\sigma$ clipping algorithm described in Section \ref{sec:specfit}. The spectra were then normalized to the median flux value in the rest-frame wavelength range 1425--1475 \angstrom and projected onto a common frame (with a pixel size of 0.5 \angstrom). We then found the median flux value from all of the quasars in the subsample to determine each pixel in the median stacked spectrum. In total, there were 19 X-ray strong, 597 X-ray normal, and 40 X-ray weak typical quasars, and 39 WLQs, used to generate the four stacked spectra shown in panel (a) of Figure~\ref{fig:linestack}. As expected from Table~\ref{tab:lumadj_civ}, the median stacked \civ emission line for the X-ray strong typical quasars is much stronger, and has a smaller blueshift, than for the X-ray normal and X-ray weak typical quasars. The WLQs, which are individually depicted in panel (b) of Figure \ref{fig:linestack}, exhibit a much larger blueshift than the typical quasars, and have a weaker, more asymmetric, \civ profile. Panel~(c) of Figure \ref{fig:linestack} depicts the ratio of the \mbox{X-ray} strong to X-ray normal typical quasars and X-ray weak to X-ray normal typical quasars, which illustrates that the \mbox{X-ray} weak and X-ray normal subsamples have similar \civ line profiles, whereas the X-ray strong quasars have much stronger \civ emission (as reported in Table \ref{tab:lumadj_civ}). While a detailed analysis of the shapes of the \civ line-profiles is beyond the scope of this work, we performed a basic measure of the kurtosis of the \civ emission line for the \mbox{X-ray} strong, \mbox{X-ray} normal, and \mbox{X-ray} weak typical quasar subsamples. We found that all three stacked subsamples display leptokurtic (i.e.\ have a positive kurtosis value) \civ emission-line profiles indicating that they are generally more ``peaky'' than an ordinary gaussian profile (e.g.\ \citealt{Denney2012}). The \civ line profile for the X-ray strong subsample has the largest kurtosis value ($k = 9.29 \pm 3.10$) and therefore is more strongly peaked compared to the X-ray normal ($k = 4.80 \pm 0.25$) and X-ray weak ($k = 4.48 \pm 1.11$) subsamples. Errors on the kurtosis values for each X-ray subsample were computed using a bootstrap method in which we created 500 median stacked line profiles and computed the standard deviation of the kurtosis values of all 500 profiles. 


The X-ray weak typical quasars tend to have weaker \civew and larger \civ blueshift than their X-ray strong counterparts, and seem to occupy the space between the X-ray strong typical quasars and the X-ray weak WLQs in Figure \ref{fig:civew_blueshift}. We therefore investigated whether the X-ray weak typical quasars and the X-ray weak WLQs could be connected using their luminosity-adjusted \civew and \civ blueshift values. A Spearman rank-order test on the X-ray weak typical quasars indicates, however, that there is no significant correlation between the \civew and \civ blueshift ($P_{\rm null} = 0.250$); therefore, no relationship can be extrapolated to the X-ray weak WLQs. A larger sample of X-ray weak typical quasars could aid determining if a correlation exists between \civew and \civ blueshift, and if there is a connection between X-ray weak typical quasars and X-ray weak WLQs; however, such a sample would have to be assembled in an unbiased manner. Further investigation is required to understand better the relationship between \daox, \civew, and \civ blueshift in typical quasars, and if these parameters can be related to those measured for WLQs.

\begin{table}
\centering
\caption{X-ray subset sample statistics for typical quasars}
\label{tab:lumadj_civ}

\begin{tabular}{l c r r r r}
\hline
\thead{X-ray} & \thead{\civ}  & \thead{Median} & \thead{25\%} & \thead{75\%} & \thead{IQR}  \\
\thead{Subset} & \thead{Parameter} & \thead{Value}  &  & &   \\
\thead{(1)} & \thead{(2)} & \thead{(3)} & \thead{(4)} & \thead{(5)} & \thead{(6)}  \\
\hline
 \multicolumn{6}{c}{\civew--\civ blueshift statistics} \\
 \hline
Strong & Blueshift & $172.34$ &  $-18.88$ & $490.17$ & 509.05  \\
Strong& EW & 97.08 &  59.77 & 111.92 & 52.15  \\
Normal  & Blueshift & $427.08$ &  $172.58$ & 842.99 & 670.41   \\
Normal & EW & $47.62$ &  $36.03$ & 63.87 & 27.84   \\
Weak  & Blueshift & $486.08$ &  $201.94$ & 662.16 & 460.22   \\
Weak & EW & $45.68$ &  $34.91$ & 62.04 & 27.13  \\
\hline
 \multicolumn{6}{c}{\mbox{$\Delta{\rm{log}}_{10}$(\civew)}-- \mbox{$\Delta$(\civ blueshift)} statistics} \\
 \hline
Strong & Blueshift & $-227.22$ &  $-398.71$ & $-121.92$ & 276.79  \\
Strong& log$_{10}$(EW) & 0.202 &  0.072 & 0.368 & 0.296  \\
Normal  & Blueshift & $-5.67$ &  $-253.38$ & 302.54 & 555.92   \\
Normal & log$_{10}$(EW) & $-0.008$ &  $-0.120$ & 0.103 & 0.222   \\
Weak  & Blueshift & $-147.82$ &  $-268.88$ & 248.34 & 517.22   \\
Weak & log$_{10}$(EW) & $-0.03$ &  $-0.113$ & 0.080 & 0.193  \\
\hline
\end{tabular}

\begin{flushleft}
\footnotesize{{\it Notes:} The X-ray strong (\daox $\geq 0.2$), X-ray normal \mbox{($-0.2<$\daox$<0.2$)}, and X-ray weak (\daox $\leq -0.2$) sample statistics for \civew and \civ blueshift (top six rows), and \mbox{$\Delta$(\civ blueshift)}, and \mbox{$\Delta{\rm{log}}_{10}$(\civew)} (bottom six rows). Columns 1 and 2 define the X-ray sample and the \civ blueshift ($\rm km\ s^{-1}$) or EW (\angstrom) and columns 3--6 report the median value, 25th and 75th percentile values, and interquartile range (IQR) of the data in these subsamples. The X-ray strong subsample generally has smaller blueshift values and larger \civew than the X-ray weak subsample.}

\end{flushleft}
\end{table}

\subsection{\ion{Mg}{\uppercase{ii}} relationship with \aox and \daox}

\begin{figure}
	\includegraphics[width=\columnwidth]{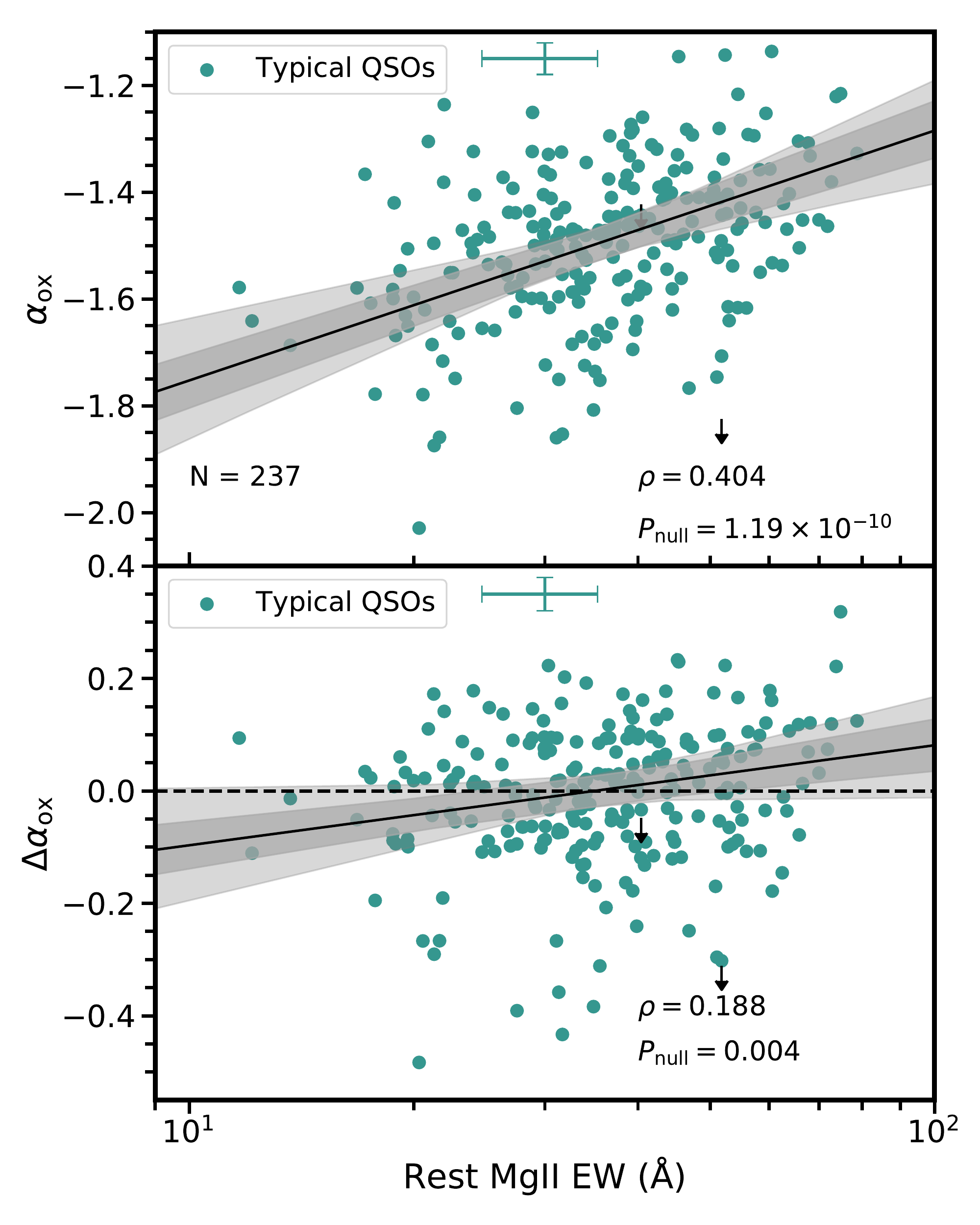}
    \caption{\aox (top panel) and \daox (bottom panel) as a function of the \mgiiew for typical quasars (green points; black arrows depict quasars with X-ray upper limits). Median errors are indicated by the green error bars at the top of each panel and the Spearman rank-order test results are reported at the bottom right, as in Figures \ref{fig:daox_civew} and \ref{fig:aoxdaox_blueshift}. We use the Bayesian fitting method of \citet{Kelly2007} to find the best-fit relation (solid black line) and the $1\sigma$ (dark grey region) and $3\sigma$ (grey region) CI. Again, \aox--\mgiiew shows a clear positive correlation within the $3\sigma$ CI, whereas the fitted correlation in \daox--\mgiiew is more uncertain. Such uncertainty could be due to a smaller sample size with decreased dynamic fitting range in \mgiiew compared to that of \civew.}
    \label{fig:aoxdaox_mgiiew}
\end{figure}

As reported in Section \ref{sec:specfit}, the redshift range of our sample provided access to the \mgii broad emission line for a subset of the quasars in our Sensitive sample. Using this \mgii subsample, we investigated the relationship between X-ray emission and the low-ionization line region in quasars, which is likely associated with emission from the quasar accretion disk (e.g.\ \citealt{Collin1987, Strateva2003, Leighly2004, Leighly2007, Shen2014}). As in Sections \ref{sec:daox_civew} and \ref{sec:daox_civbl}, we investigated whether a relationship exists between \aox, \daox, and \mgiiew. We performed a Spearman rank-order test on the quasars in our \mgii subsample which finds correlations between \aox--\mgiiew and \daox--\mgiiew (see Table \ref{tab:twoSamplestats}). Using the fitting algorithm of \citet{Kelly2007}, the best-fit relationships are
\begin{equation} 
\begin{split}
\alpha_{\rm{ox}} = (0.468 \pm 0.062)\rm{log_{10}(\mgii \ EW)} - (2.22 \pm 0.097), \\ \sigma_{\epsilon} = 0.11,
\end{split}
\label{eq:aox_mgiiew}
\end{equation}

\begin{equation} 
\begin{split}
\Delta\alpha_{\rm{ox}} = (0.177 \pm 0.056)\rm{log_{10}(\mgii \ EW)} - (0.273 \pm 0.087), \\ \sigma_{\epsilon} = 0.10.
\end{split}
\label{eq:daox_mgiiew}
\end{equation}
These fits, as well as the $1\sigma$ and 3$\sigma$ CIs, are presented in Figure~\ref{fig:aoxdaox_mgiiew}. We find that \aox is positively correlated with \mgiiew, but the trend in \daox--\mgiiew becomes flatter. The uncertainty in the \daox--\mgiiew relationship is significant due to a large amount of scatter in \daox and a lack of dynamic range of \mgiiew.

\begin{figure*}
	\includegraphics[width=0.9\textwidth]{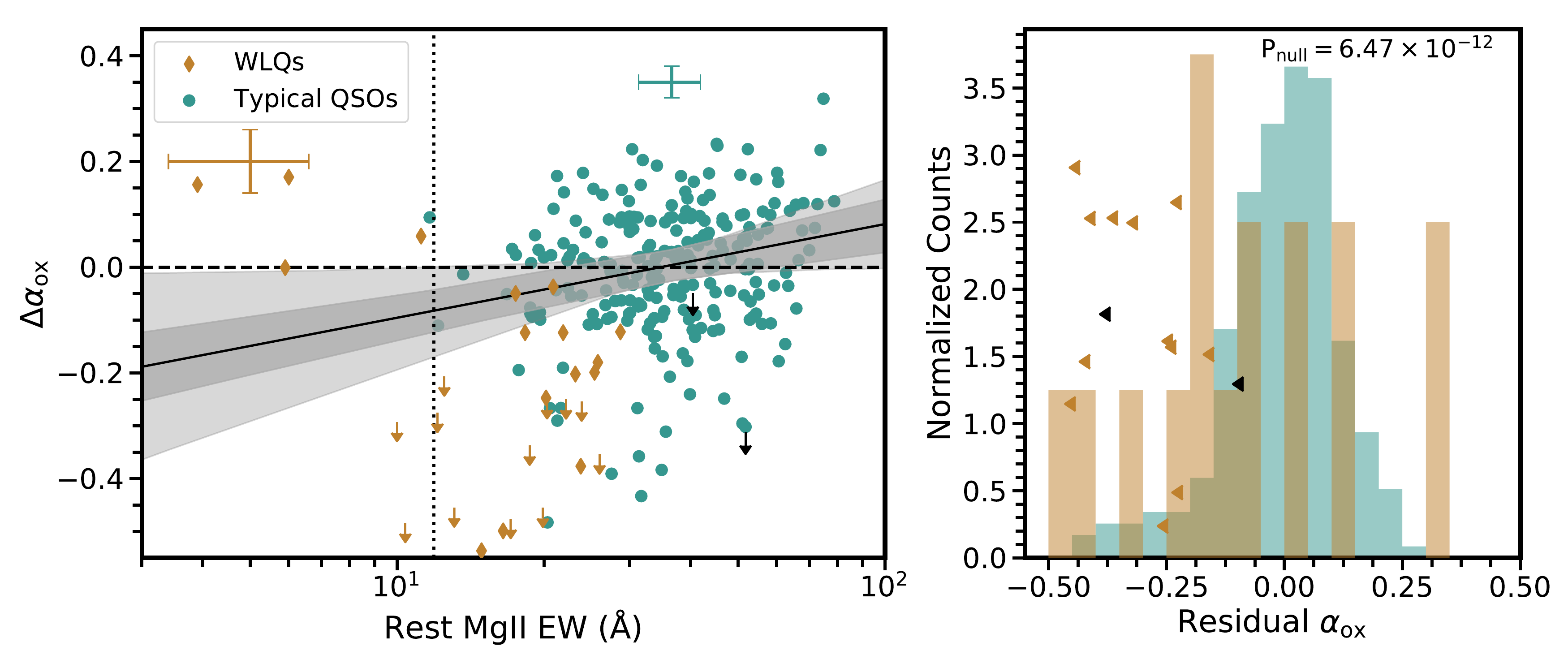}
    \caption{Left: The same as the bottom panel of Figure \ref{fig:aoxdaox_mgiiew} with the WLQs (brown diamonds/arrows) included, akin to Figure 5 of \citet{Wu2012} and Figure~4 of \citet{Plotkin2015}. The vertical dotted line depicts the location of the \mgiiew 3$\sigma$ outliers from Figure 3 of \citet{Wu2012}, which shows that many of the WLQs (defined by \civew) have weak, but not exceptionally weak, \mgiiew. Median errors are indicated for the typical quasars and WLQs by the green and brown error bars at the bottom of the plot, respectively. The \daox--\mgiiew correlation from Equation \ref{eq:daox_mgiiew} is used to compute the residual \daox values, as before. Right: Distributions of the residual \daox for both the typical quasars and WLQs using the best-fit trend in the typical quasars. A Peto-Prentice comparison of these two distributions reports, with high probability, that they are not the same.}
    \label{fig:daox_mgiiew}
\end{figure*}

The \mgiiew distribution for the WLQs is depicted in the left panel of Figure \ref{fig:daox_mgiiew} along with that for the typical quasars. In general, the WLQs have smaller values of \mgiiew than the typical quasars in our sample; these values, however, are not especially small when compared to the striking difference in \civew values between the WLQs and typical quasars (compare Figures \ref{fig:resid_med} and \ref{fig:daox_mgiiew}). Moreover, the location of the WLQs in this space does not follow the expected relation, even within the large uncertainty on Equation~\ref{eq:daox_mgiiew}. This difference is illustrated in the right-hand panel of Figure~\ref{fig:daox_mgiiew} which depicts the residual \daox for both the WLQs and typical quasars from the best-fit relationship. A Peto-Prentice test confirms that the WLQs and the typical quasars do not have similar residual \daox distributions ($P_{\rm null}= 6.47 \times 10^{-12}$). As with the \civ parameters, we examined how a luminosity dependence of the \mgiiew might affect this result; we find no difference if we adjust \mgiiew for luminosity. The \mgii emission line is a good redshift indicator \citep{Shen2016}, and is used as either a direct measure of the redshift, or as a metric to calibrate the algorithms that estimate the redshift (e.g.\ \citealt{Paris2017}) of the quasars in our samples; therefore, we do not investigate the relationship between \aox, \daox, and \mgii blueshift  (e.g.\ \citealt{Plotkin2015}). Significant blueshifts ($>500\ \rm km\ s^{-1}$) of the \mgii emission line were only found in the WLQs with the highest \civ blueshifts in \citet{Plotkin2015}, and are expected to be small, if present, in typical quasars. Finally, we find a correlation between \aox (\daox) and \mgii FWHM with a Spearman rank-order $\rho=0.25$ ($\rho=0.19$) and a null probability of $P_{\rm null}= 1 \times 10^{-4}$ ($P_{\rm null}= 6 \times 10^{-3}$); however, given the complexity in fitting the \mgii emission line, and the lack of \mgii FWHM measurements for the WLQs, further analysis of this correlation is beyond the scope of this work.

\begin{figure}
	\includegraphics[width=\columnwidth]{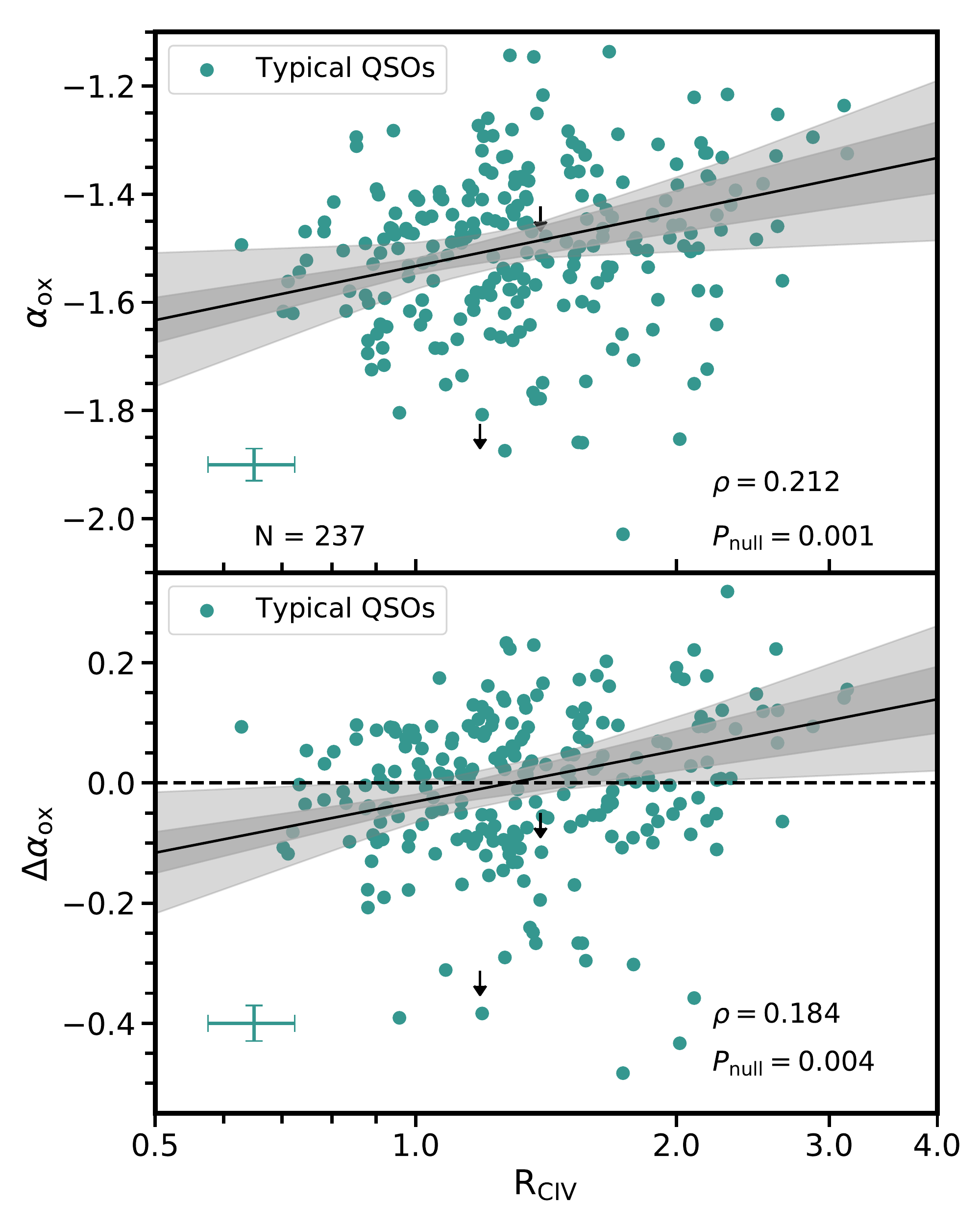}
    \caption{\aox (top panel) and \daox (bottom panel) as a function of the relative strengths of the high- and low-ionization lines, quantified by the ratio of \civew to \mgiiew (R$_{\rm \civ}$), of the typical quasars (green points; black arrows indicate quasars with X-ray upper limits). The median errors are depicted as green error bars at the bottom left of each panel, and the Spearman rank-order test results are reported in the bottom right. Using the algorithm of \citet{Kelly2007}, we fit a trend to the typical quasars (solid black line) in each panel, and the black dashed horizontal line in the bottom panel indicates the expected X-ray strength given $L_{2500}$. The $1\sigma$ and $3\sigma$ CI are shown as dark grey and grey regions, respectively. Unlike the previous analyses, the slope of the \aox--R$_{\rm \civ}$ relationship and the \daox--R$_{\rm \civ}$ relationship are similar.}
    \label{fig:aoxdaox_Rciv}
\end{figure}

\begin{figure*}
	\includegraphics[width=0.9\textwidth]{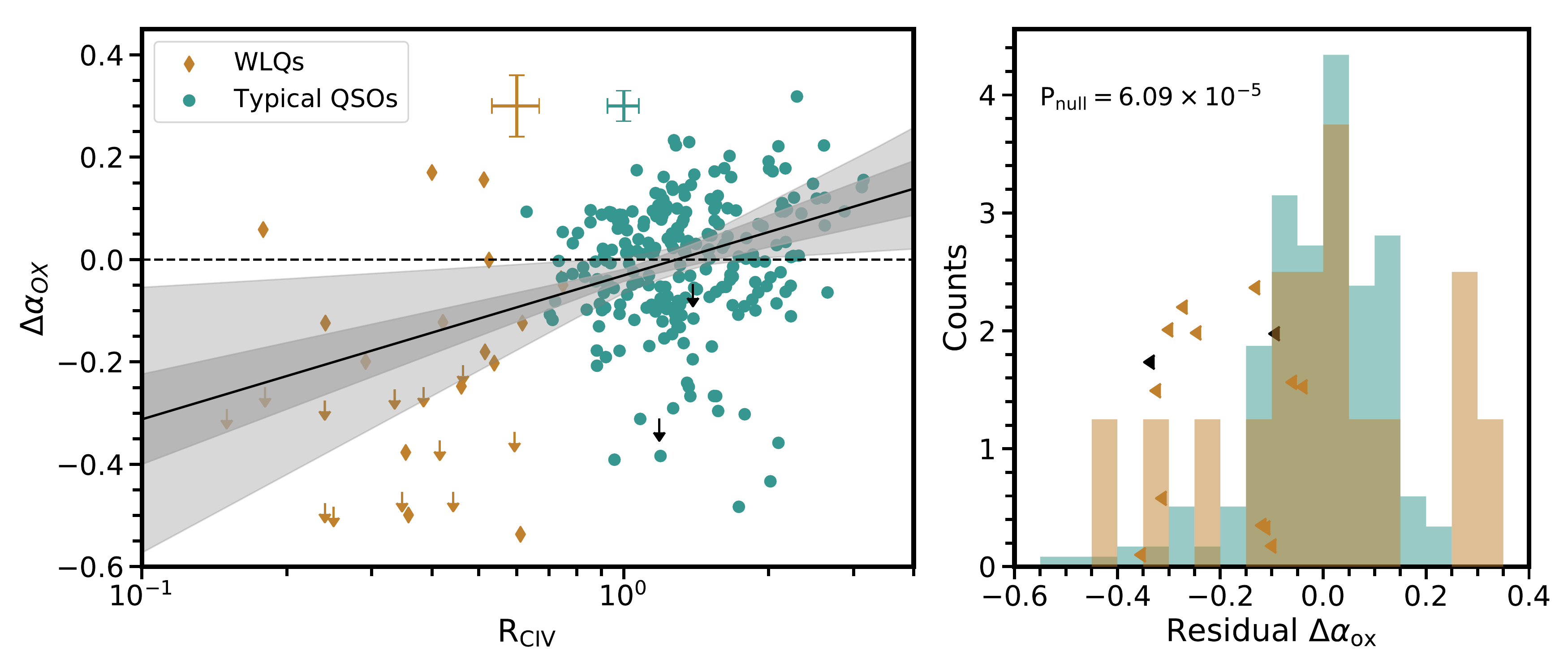}
    \caption{Left: We include the WLQs (brown diamonds; brown arrows depict WLQs with X-ray upper limits) to the bottom panel of Figure \ref{fig:aoxdaox_Rciv}. It is clear that the WLQs occupy lower values of R$_{\rm \civ}$ than the typical quasars. The median errors are depicted as green and brown error bars at the top for the typical quasars and WLQs, respectively. Right: As before, we estimate the residual \daox by examining the difference between the measured \daox and the best-fit trend line for both the typical quasars and WLQs. A Peto-Prentice test demonstrates that the typical quasars and WLQs are not drawn from the same distribution; however, $P_{\rm null}$ is larger than in the \mgiiew relationship, indicating a lower significance at which we can reject the null hypothesis. We cannot predict the X-ray weakness of the WLQs using the ratio of the high- to low-ionization lines in the typical quasars.}
    \label{fig:daox_Rciv}
\end{figure*}

Finally, we investigated the relationship between the X-ray emission and the relative strengths of the high- and low-ionization lines of quasars \mbox{\citep{Plotkin2015, Shemmer2015}}. We computed the parameter R$_{\civ} = {\rm{log}}_{10}$(\civew / \mgiiew) to use as a metric for the difference in relative emission-line strengths\footnote{\citet{Plotkin2015} computed a similar parameter using \ion{H}{$\beta$} rather than \mgii; however they did not utilize the X-ray information for their quasar sample.} for the quasars in our \mgii subsample and the WLQs. Again, a Spearman rank-order test indicates that both \aox and \daox are correlated with R$_{\civ}$ (see Table \ref{tab:twoSamplestats}). The top panel of Figure \ref{fig:aoxdaox_Rciv} presents \aox as a function of R$_{\civ}$, and the bottom panel depicts \daox as a function of R$_{\civ}$. We find the best-fit relationships to be
\begin{equation} 
\begin{split}
\alpha_{\rm{ox}} = (0.332 \pm 0.095)\rm{log_{10}(\rm{R_{\civ}})} - (1.533 \pm 0.014), \\ \sigma_{\epsilon} = 0.13,
\end{split}
\label{eq:aox_Rciv}
\end{equation}

\begin{equation} 
\begin{split}
\Delta\alpha_{\rm{ox}} = (0.284 \pm 0.077)\rm{log_{10}(\rm{R_{\civ}})} - (0.031 \pm 0.012), \\ \sigma_{\epsilon} = 0.10,
\end{split}
\label{eq:daox_Rciv}
\end{equation}
 which was again fitted using the algorithm of \citet{Kelly2007} for the \mgii subsample. Following the previous procedure, we extrapolate Equation \ref{eq:daox_Rciv} to small values of R$_{\civ}$ (see the left panel of Figure~\ref{fig:daox_Rciv}) to find the expected \daox for the WLQ sample. The residual \daox was calculated using Equation~\ref{eq:daox_Rciv} (shown in the right panel of Figure~\ref{fig:daox_Rciv}), and a Peto-Prentice test was applied to the distributions of the typical quasars and the WLQs. While the two sets are still not drawn from the same distribution, we reject the null hypothesis of similarity with lower significance ($P_{\rm null}=6.09\times10^{-5}$; see Table~\ref{tab:twoSamplestats}) than for the \daox--\mgiiew relationship. An increased sample size over a larger dynamic range is needed to reduce further the confidence interval on this correlation, and to investigate whether the X-ray weakness of the WLQs can be better predicted using R$_{\civ}$ compared to \civew for the typical quasars.

\section{Discussion}
We have compiled a large, unbiased sample of radio-quiet quasars that have both X-ray and rest-frame UV broad emission-line measurements, greatly increasing the sample size relative to Sample B of \citet{Gibson2008}. Using the quasars in our sample, we investigated the relationships between X-ray strength and different parameters of the rest-frame UV broad emission lines. Many of these correlations have been or can be explained in the context of the ``disk-wind'' model for typical quasars (e.g.\ \citealt{Murray1995, Elvis2000, Leighly2004, Richards2011}). This model splits the quasar broad line region into two parts: the ``disk'' component is composed of gas near the disk in orbit around the central black hole, and the ``wind'' component exhibits both an outflow velocity and an orbital velocity (e.g.\ \citealt{Collin2006}). In the standard picture, much of the gas in the wind region is closer to the central source, and therefore interacts directly with the ionizing radiation from the central continuum, whereas the gas in the disk region is further from the central engine, and is exposed to a continuum filtered by the wind \mbox{(e.g.\ \citealt{Leighly2004, Leighly2007, Richards2011})}. The high-ionization emission lines are therefore thought to be in the wind region (particularly in high-luminosity quasars), since it interacts directly with the ionizing continuum, whereas the low-ionization emission lines are believed to originate in the disk region. We adopt \aox as a basic measure of the hardness of the ionizing SED, where the amount of ionizing radiation increases with the hardness of the SED (e.g.\ larger values of \aox). We discuss the measured correlations between the rest-frame UV emission lines and X-ray strength of our quasars in the context of this model.

\subsection{\aox and \daox correlations among typical quasars}\label{sec:qso_corr}

In our large set of unbiased quasars, the abundance of \civ (\civew) is positively correlated with both \aox and \daox, which is consistent with previous results (e.g.\ \citealt{Green1998, Gibson2008}). Adopting the parlance of the disk-wind scenario, the significant correlation found between \aox and \civew suggests that quasars with harder SEDs generally produce larger amounts of high-ionization gas. Moreover, the correlation between \daox and \civew demonstrates that this interpretation holds even after normalizing for quasar luminosity. Our analysis also demonstrates with high confidence that the \daox--\civew correlation exists (though is weaker than the \aox--\civ EW correlation), and is not due to statistical noise, as a result of a small sample size, or an artifact related to sample bias. 

In the disk-wind model, the high-ionization lines display both an orbital component of velocity along with an outflowing component, where we use the blueshift of the high-ionization broad emission lines as a basic proxy for the velocity of the outflow. Our analysis of \aox and \civ blueshift demonstrates that there is a relationship between these two parameters. We confirm that a correlation also exists when we account for the luminosity dependence of \aox; however, the relationship between \daox and \civ blueshift is less certain. While we expect the \civ emission line to display a blueshifted component, the blueshift should not necessarily be linearly related to the quasar SED; rather, there is likely a range of parameters that balance the ionizing radiation and the line-driving radiation. For example, if the broad line region is oversaturated with ionizing photons, the gas becomes over-ionized, and the line-driving by the less-energetic UV radiation will be inefficient (e.g.\ \citealt{Murray1995, Elvis2000, Leighly2004, Richards2011, Plotkin2015}). Such a phenomenon is present in our analysis, where quasars that are X-ray strong typically have smaller blueshift values than the X-ray weak quasars. Additionally, orientation effects could complicate the relationship between the measured \aox and outflow velocity (the top panel of Figure \ref{fig:aoxdaox_blueshift}). The large scatter seen in the bottom panel of Figure \ref{fig:aoxdaox_blueshift} also suggests that there is more complex physics and orientation effects at work than can be modeled by a linear fit. We additionally investigated \aox (and \daox) vs. \mbox{log$_{10}$(\civ blueshift)} and find that a simple power-law model cannot accurately capture the complex physics that governs the blueshift either.

Joint analyses of quasar \civew and blueshift properties have also provided a compelling link to the hardness of the SED, such that objects with both large \civew \emph{and} small blueshift have a harder SED than objects with both smaller \civew and larger blueshift (e.g.\ \citealt{Leighly2004,Richards2011,Kruczek2011,Wu2011,Wu2012,Shen2013}). We find, as in \mbox{\citet{Wu2011,Wu2012}}, that many of the X-ray strong (\daox $\geq 0.2$) quasars occupy a slightly different region in this parameter space than the X-ray weak \mbox{(\daox $\leq -0.2$)} quasars. The X-ray normal (\daox $\approx 0$) quasars, however, display a larger range in \civ blueshift compared to the other subsets (see Table \ref{tab:lumadj_civ}). This scatter could be attributed to the X-ray normal quasars hosting a wider range of SEDs than the X-ray strong and X-ray weak quasars which, in turn, produces disk and wind systems with more diverse properties. An in-depth analysis of the distribution of SEDs for these quasars (similar to \mbox{\citealt{Kruczek2011}}) would provide more information about the trade-off between the disk and wind components.

The size of our unbiased sample also enables an investigation of the relationships between the low-ionization \mgii emission line and \aox (and \daox). In the disk-wind model, the low-ionization emission-line region lies close to the disk and is subject to a modified SED that has been filtered by the high-ionization wind (e.g.\ \citealt{Leighly2004,Richards2011}). Since the relative number of ionizing photons generally increases with \aox, we might expect more photons to be filtered through the wind component, and thus the filtered continuum can ionize more gas in this low-ionization region. Our investigation found that the \mgiiew is positively correlated with \aox and \daox, which is consistent with this picture. While we attribute the uncertainty in slope to sample statistics, the measured continuum may not necessarily be directly related to the continuum seen by the low-ionization region. The filtered continuum is dependent on the properties of the wind (e.g.\ ionization parameter) which could complicate the trend we found between \aox (and \daox) and \mgiiew. Interestingly, however, the Spearman rank-order correlation statistics are similar between the \aox--\civew and \aox--\mgiiew relations (see Table~\ref{tab:twoSamplestats}), which is consistent with our interpretation that more ionizing photons are passing through the wind gas and ionizing the disk region.
 
We investigated \aox, \daox, and the relative strengths of the high- and low-ionization lines using the ratio of \civew to \mgiiew. Such a comparison potentially shows the mutual reaction of the wind and disk components to different ionizing continua. There is a positive correlation between the strength of the ionizing radiation and the relative strength of the \civ emission line. In the context of the disk-wind model, an absence of correlation in this space would imply that the shape of the SED (hard or soft) influences the high- and low-ionization lines to the same degree. The positively correlated relationship, however, indicates that a hard SED tends to create a more disk-like system, whereas a soft SED generates a system with large outflows (wind-like). Again, the scatter present in this relation can be attributed to slight differences in the SED of each quasar.

Finally, there is a small subset of twelve quasars that seem to be inconsistent with the most-straightforward disk-wind model predictions. These quasars have high \civew \mbox{($\geq60$ \angstrom)} and are measured to be X-ray weak \mbox{(\daox$\leq-0.2$)} as is apparent in Figure~\ref{fig:daox_civew}. The average \mbox{X-ray} power-law photon index of these twelve quasars is very hard \mbox{($\langle \Gamma_{\rm eff} \rangle = 0.85^{+0.3}_{-0.1}$)}, which is indicative of soft X-ray absorption like that often seen in BAL quasars (e.g.\ \citealt{Brandt2000}). We have, however, systematically and visually\footnote{One of these twelve objects, SDSS--J$231427.50-025251.1$, is flagged as a mini-BAL quasar.} identified and removed BAL quasars from our sample, making such X-ray weakness puzzling. A plausible explanation could be that these quasars were BAL quasars during the X-ray observation, but there was no BAL at the time the SDSS spectrum was measured (e.g.\ \citealt{FilizAk2012,Sameer2019}); however, such instances of BAL disappearance are rare ($\approx$ 5\% of BAL quasars; \citealt{DeCicco2018}). It is also unlikely that extreme X-ray variability is causing the large degree of X-ray weakness in these objects. To overlap with the other high \civew quasars, these twelve quasars would need to display a $\approx$ 100--200\% increase in soft X-ray flux, which corresponds to an increase in \daox by $\approx$ 0.1--0.2. Such extreme variability is rare at high luminosities (e.g.\ see \citealt{Gibson2012}). Further investigation of these quasars could provide additional insights into quasar environments.

\subsection{X-ray connections between typical quasars and WLQs}\label{sec:wlqqso_corr}

Based on Figure \ref{fig:Mi_z}, the typical quasars in this investigation clearly span a different range in luminosity than the WLQs from \citet{Ni2018}, and therefore a direct comparison of the \aox values is inappropriate. Instead, we only compare their measured \daox values to remove potential biases due to the luminosity dependence of \aox. The WLQs display a larger scatter in \daox than the typical quasars, generally have much larger \civ blueshifts, and tend to have higher Eddington ratios \citep{Luo2015, Marlar2018}. To explain such behavior, we consider the ``shielding'' model \citep{Luo2015, Ni2018}, which proposes that a geometrically thick inner accretion disk\footnote{Such a disk geometry can occur for quasar systems accreting at high Eddington ratio (see \citealt{Ni2018} and references therein).} blocks the X-ray and extreme-UV emission from reaching the broad emission-line region. The impinging SED of the ionizing continuum then becomes much softer, causing the \civ emission line to be exceptionally weak. Furthermore, depending on the observer's line-of-sight, the observed X-ray emission could be extremely weak if the disk blocks the X-ray photons. Our analysis of \daox and \civew is consistent with this scenario; the WLQs are more scattered around the best-fit relationship (derived using the typical quasar population) than the intrinsic scatter seen for the typical quasars (see Figure \ref{fig:resid_med}). In general, the WLQs tend to be X-ray weaker than expected from the \daox--\civew trend for the typical quasars \citep{Wu2011,Wu2012,Luo2015,Ni2018}; however, there are a few notable WLQs that are much X-ray stronger than expected (though are still X-ray normal; see Figure \ref{fig:daox_civew}). To achieve nominal X-ray emission, the ``shielding'' model predicts that these WLQs are observed at high inclination angles off of the disk axis.

Furthermore, the WLQs display \civ blueshifts much larger than those of the typical quasars (e.g.\ see Figure 12 of \citealt{Ni2018}). As with \civew, many of the WLQs are X-ray weaker than expected based upon a linear extrapolation of the relationship between \daox and the \civ blueshifts for the typical quasars (see Figure \ref{fig:daox_blueshift}); however, a linear relationship is not likely to be the best indicator of physical behavior in this space \citep{Wu2011}. While our enlarged sample of typical quasars does span a wider range than those in \citet{Gibson2008}, and seems to extend toward the WLQs (particularly for the X-ray weak, high blueshift typical quasars), we did not discover any direct link between the X-ray weak typical quasars and WLQs. A larger sample of WLQs would be beneficial to further fill this space.

The WLQs tend to have smaller \mgiiew than the typical quasars (also see \citealt{Wu2012}); however, they are generally not extreme outliers as in the case for the \civew. We find that the \mbox{X-ray} emission of the few WLQ that have extremely small \mgiiew are not well predicted by the extrapolated \daox--\mgiiew relation from the typical quasars and, interestingly, are generally X-ray normal. In the ``shielding'' model, the higher-energy extreme-UV photons that are able to produce \civ (at $\approx 47.9$ eV) are presumed to be partially or mostly obscured (resulting in the small \civew); however, the amount of far-UV radiation responsible for driving the wind and ionizing \mgii (at \mbox{$\approx 7.6$ eV}) could be only somewhat smaller than that of the typical quasars. Such behavior could explain why the \mgiiew is not excessively weak and why the WLQs exhibit higher-velocity outflows than the typical quasars. Furthermore, it has been shown that the IR-to-near-UV SEDs of WLQs (e.g.\ Figure~7 of \citealt{Luo2015}) are similar to those of typical quasars, so the expectation of approximately nominal levels of \mbox{$\approx 7.6$ eV} photons is not unreasonable.

Our analysis of the \daox--R$_{\rm \civ}$ relationship is also consistent with the ``shielding'' model, where the ratio of high- and low-ionization line EWs is a proxy for the shape of the UV continuum that interacts with the entire broad line region. The correlation between \daox and R$_{\rm \civ}$ for the typical quasars does not predict the X-ray weakness of the WLQs. Moreover, R$_{\rm \civ}$ for the WLQs tends to be smaller than for the typical quasars. Early models of WLQs supposed that WLQs had ``anemic'' broad emission-line regions (e.g.\ \citealt{Shemmer2010,Niko2012}). This idea, in its simplest form, would require the broad emission-line region to be gas deficient. If this were the case, however, we would likely expect R$_{\rm \civ}$ of the WLQs and the typical quasar population to be similar. Since the WLQs and typical quasars have different R$_{\rm \civ}$ values, we can rule out this simple ``anemic'' broad emission-line region model (see also \citealt{Plotkin2015}). 

Overall, we found that the WLQs generally do not lie on an extrapolation of any of the relations found for typical quasars in this work. Our work suggests that the broad emission-line region in WLQs could be subject to a softer ionizing SED than the typical quasars. We also find no apparent relation between the X-ray weak WLQs and X-ray weak typical quasars, despite having similar {\emph{observed}} X-ray emission in terms of \daox. These results are consistent with the ``shielding'' model, which proposes that the inner accretion disk of the WLQs becomes thick due to their high Eddington ratios. The thick disk blocks the higher-energy ionizing UV and X-ray photons that are responsible for exciting the gas in the high-ionization region, resulting in differing \civ abundances between the typical quasars and WLQs. Testing the dependence on Eddington ratio will require more precise emission-line measurements, particularly of the \ion{H}{$\beta$} emission line, for each of our typical quasars.


\section{Summary and Future Investigations}

This investigation has greatly increased the sample size of unbiased SDSS quasars with X-ray coverage relative to \citet{Gibson2008}. Using this large sample, we more tightly constrained correlations between X-ray and optical/UV broad emission-line properties in typical (radio-quiet, non-BAL) quasars, and we compared these properties to those of WLQs, using the WLQ sample compiled in \citet{Ni2018}. We assembled a large sample of 2106 quasars from the full SDSS quasar archive (through DR14) that were serendipitously observed with \Chandra and are within the redshift range $1.7 \leq z \leq 2.7$. The sample was further reduced to 753 quasars by imposing a cut on exposure time and the off-axis angle of the \Chandra observation, which produced our Sensitive sample of quasars with high detection fraction. We fit the continuum, \civ emission line, and \mgii emission line for each SDSS spectrum of these quasars using the PyQSOFit software. We measured the correlations between the X-ray, \civ, and \mgii emission-line properties for our data set, and extrapolated those relationships for comparison to the WLQ sample. These results were interpreted in the context of the disk-wind model for typical quasars and the ``shielding'' model for WLQs. The main results from this paper are the following:

\begin{enumerate}

\item Our Full sample of quasars consists of a larger number of serendipitous X-ray observations (Figure \ref{fig:off-axis_hist}) and spans a much wider range in luminosity than the \citet{Gibson2008} \mbox{Sample B} \mbox{(Figure \ref{fig:Mi_z})}. Using our Sensitive sample (Figure \ref{fig:exposure_offaxis}), we found the best-fit relationship between \aox and ${\rm{log}}_{10}(L_{2500})$, and calculated \daox using this relationship for both our quasars (Figure \ref{fig:aox_logL2500}) and the WLQs in \citet{Ni2018}.

\item Comparing \aox and \daox with \civew in Figure \ref{fig:daox_civew} reveals a clear positive correlation. The best-fit \daox--\civew relationship in our sample does not differ largely from that in \citet{Gibson2008}; however, we are able to constrain better the slope. This trend is consistent with the disk-wind model for typical quasars. Computing the residual \daox from our best-fit trend demonstrates that the WLQs are not drawn from the same distribution as the typical quasars in this space (Figure \ref{fig:resid_med}); they tend to be much \mbox{X-ray} weaker than expected; however, there are a few X-ray normal WLQs. This result is consistent with previous results \citep{Wu2011,Wu2012,Plotkin2015,Luo2015,Ni2018}, and can be understood by invoking the ``shielding" model for WLQs \citep{Luo2015}.

\item We sample a much larger blueshift space than previous investigations and find a significant correlation between \aox and \civ blueshift using a Spearman rank-order test (see Figure \ref{fig:aoxdaox_blueshift}). Furthermore, we extrapolate the \daox--\civ blueshift trend to large \civ blueshifts which reveals that the WLQs tend to be X-ray weaker than predicted from the trend (\mbox{Figure \ref{fig:daox_blueshift})}. 

\item We investigated the \civew--\civ blueshift parameter space (top panel of Figure \ref{fig:civ_L2500}), and the luminosity-adjusted parameters $\Delta$\civew and $\Delta$(\civ blueshift) (bottom panel of Figure \ref{fig:civew_blueshift}). The \civ emission lines of the WLQs remain much more blueshifted than those of the typical quasar population, despite adjusting for luminosity. Separating the typical quasars and WLQs into X-ray strong, X-ray normal, and X-ray weak subsamples recovers a similar SED evolution in this parameter space as previously reported (\citealt{Richards2011, Kruczek2011, Wu2011}). Furthermore, we compared the median stacked \civ emission-line profiles of these three X-ray subsamples and the WLQs (Figure \ref{fig:linestack}). We found no correlation between the X-ray weak typical quasars and the X-ray weak WLQs in this space.

\item The \mgii sample was sufficiently large (237 quasars) to investigate how \aox and \daox are related to \mgiiew (Figure~\ref{fig:aoxdaox_mgiiew}). There is a significant correlation between \aox and \mgiiew, and a slight correlation between \daox and \mgiiew. A Peto-Prentice test on the \daox residuals finds that the WLQs and the typical quasars are not drawn from the same distribution (Figure~\ref{fig:daox_mgiiew}). The \mgii EWs of the WLQs tend not to be extraordinarily weaker than those of the typical quasars (\citealt{Wu2012}; \citealt{Plotkin2015}) as is the case for \civ, which could be due to the typical quasar and WLQ low-ionization emission-line regions seeing similar SEDs out to the far-UV.

\item Finally, we investigate the relative responses of the \civew and \mgiiew with changing \aox and \daox, and find a slight positive correlation between \aox (and \daox) and $\rm R_{\civ}$ in the typical quasar population \mbox{(Figure \ref{fig:aoxdaox_Rciv})}. The residual \daox of the WLQs and typical quasars are not similarly distributed (Figure \ref{fig:daox_Rciv}); however, this correlation has much lower significance than the results of the tests between \daox and \mgiiew. The positive slope implies that the harder SEDs are more disk-dominated systems and the softer SED are more wind-dominated. The ionizing radiation is largely blocked in the WLQs, so they tend to be highly wind-dominated.  

\end{enumerate}

This work could be extended in numerous ways. First, increasing the sample size of WLQs with both X-ray and broad-line measurements would enhance our understanding of how they populate the above parameter spaces and provide a more significant comparison sample. This can be done by targeting further new WLQs from SDSS with sensitive X-ray facilities such as \Chandra. Improving the X-ray data sensitivity for the WLQs in \citet{Ni2018} that have X-ray upper limits will also provide more precise locations of WLQs in the parameter spaces in this work, and will strengthen the quality of the comparisons between typical quasars and WLQs. Moreover, repeated measurements of the same WLQs will allow us to measure the (poorly understood) variability of these objects, and to quantify how much these fluctuations affect their locations in the aforementioned parameter spaces. Furthermore, increasing the sample of typical quasars with large \civew \mbox{($\geq 120$ \angstrom)} and \mgiiew \mbox{($\geq 75$ \angstrom)} will better constrain the slopes of the correlations presented here. Additional investigation is also required to understand the complex relationships between \civew, \civ blueshift, and \daox, by performing detailed fits of the X-ray spectra, as well as investigating their dependence on the Eddington ratio. Along these lines, it would be useful to obtain measurements of the UV \ion{Fe}{ii} EW, as well as the \ion{[O}{iii]} and \ion{H}{$\beta$} emission-line properties for a subset of typical quasars and WLQs. The UV \ion{Fe}{ii} EW has been found to correlate with \daox \citep{Luo2015,Ni2018} in the WLQs, and thus it would be interesting to investigate with respect to the typical quasars as well. The \ion{[O}{iii]} and \ion{H}{$\beta$} emission-line measurements would require a significant amount of observation time using near-infrared telescopes,\footnote{Observations such as this are already underway via the GNIRS-Distant Quasar Survey (e.g.\ \citealt{Matthews2018}); see also \url{http://www.gemini.edu/node/12726}.} but would further probe the low- and intermediate-ionization regions. Measurements of \ion{[O}{iii]} would provide a more reliable determination of the systemic redshift of the quasar, which would enable examination of the \mgii blueshift correlation with \aox and \daox as suggested by \citet{Plotkin2015}. Furthermore, the \ion{H}{$\beta$} emission line could also be used to estimate physical parameters of the quasars such as black-hole mass and Eddington ratio. Finally, further investigation of the population of X-ray weak typical quasars is needed since they tend to be extreme outliers in most of our parameter spaces. This would likely require further X-ray observations to obtain quality X-ray spectra of these objects.

\section*{Acknowledgements}

We thank the referee for helping to improve this paper with their constructive feedback. We also thank Shifu Zhu and Gordon Richards for their comments and discussion throughout this work. JDT, WNB, and QN acknowledge support from NASA ADP grant 80NSSC18K0878, the \Chandra X-ray Center grant G08-19076X, the V.\ M.\ Willaman Endowment, and Penn State ACIS Instrument Team Contract SV4-74018 (issued by the \Chandra X-ray Center, which is operated by the Smithsonian Astrophysical Observatory for and on behalf of NASA under contract NAS8-03060). BL acknowledges financial support from the National Key R\&D Program of China grant 2016YFA0400702 and National Natural Science Foundation of China grant 11673010. XP acknowledges support from the Natural Science Foundation of Jiangsu Province (Grant No. BK20150870). WY thanks the support from the National Science Foundation of China (NSFC-11703076) and the West Light Foundation of the Chinese Academy of Sciences (Y6XB016001). The \Chandra ACIS team Guaranteed Time Observations (GTO) utilized were selected by the ACIS Instrument Principal Investigator, Gordon P.\ Garmire, currently of the Huntingdon Institute for X-ray Astronomy, LLC, which is under contract to the Smithsonian Astrophysical Observatory via Contract SV2-82024.


For this research, we use the Python language along with Astropy\footnote{\url{https://www.astropy.org/}} \citep{astropy2018}, Scipy\footnote{\url{https://www.scipy.org/}} \citep{scipy}, and TOPCAT\footnote{\url{http://www.star.bris.ac.uk/~mbt/topcat/}} \citep{Taylor2005}.


\appendix

\section{A sample with a higher detection fraction}\label{sec:appendix}

\begin{figure}
	\includegraphics[width=\columnwidth]{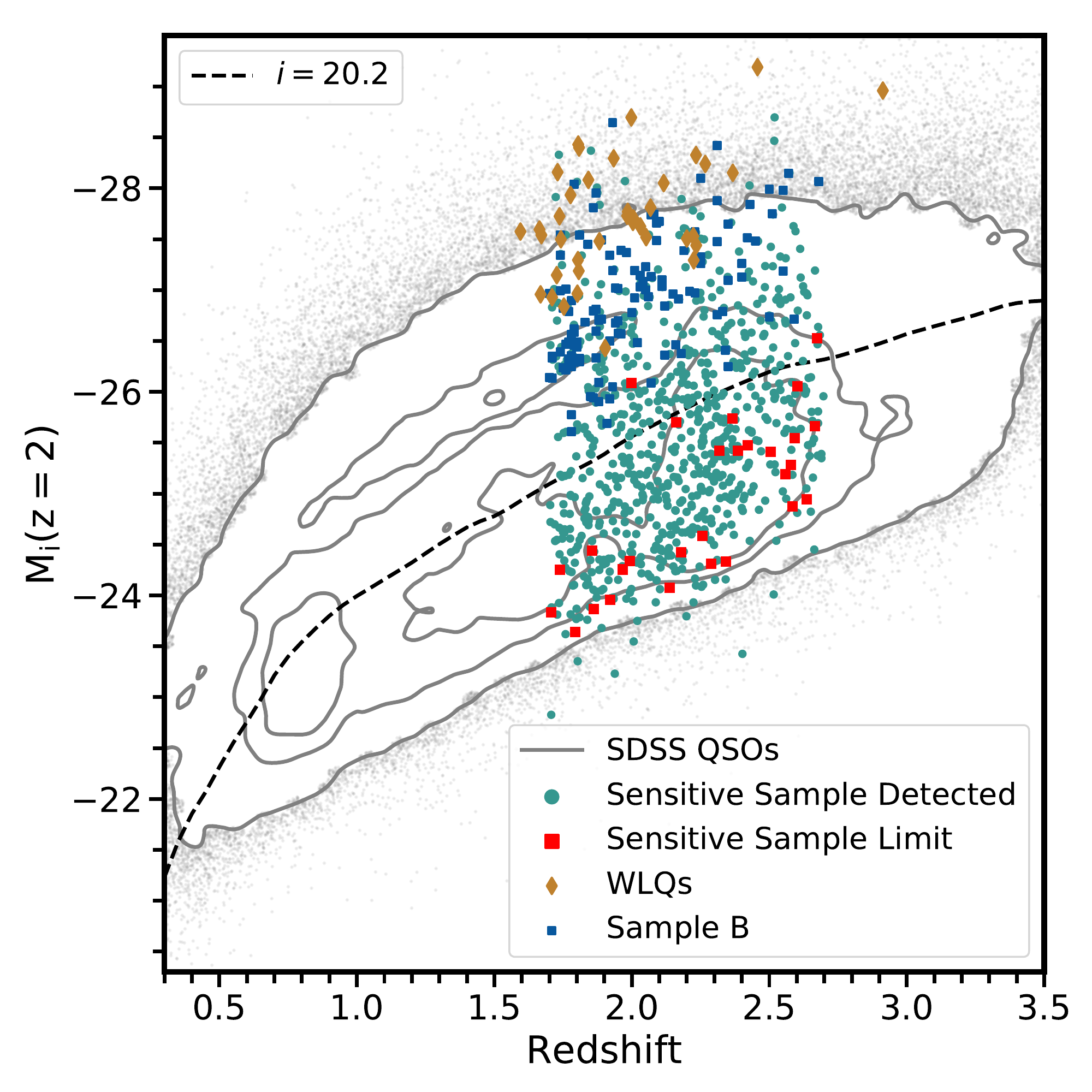}
    \caption{Similar to Figure \ref{fig:Mi_z}; however, we depict the absolute magnitude as a function of redshift for only our Sensitive sample of quasars (green points; red squares represent quasars with 90\% confidence X-ray upper limits). The dashed black curve shows the change in absolute magnitude with redshift for a quasar with a constant apparent magnitude of $i=20.2$. Restricting the Sensitive sample of quasars to only those with apparent magnitudes brighter than 20.2 ($i\leq 20.2$) generates a smaller sample than was used in our investigation (304 quasars; 302 X-ray detections, 2 \mbox{X-ray} upper limits) but has an even higher detection fraction (99.3\%). This subsample is recommended for projects that require a quasar sample with a high X-ray detection fraction, as opposed to just a large sample.}
    \label{fig:Appdx1}
\end{figure}

The 100\% X-ray detection fraction for Sample B of \citet{Gibson2008} has made it an effective comparison sample when investigating the X-ray properties of different types of quasars, and it has been extensively used over the past decade. For example, Sample B has been used to compare the X-ray emission from WLQs (e.g.\ \citealt{Shemmer2009,Wu2011,Wu2012,Luo2015,Ni2018}), BAL and mini-BAL quasars (e.g.\ \citealt{Gibson2009, Wu2010,Luo2014}), and radio-loud quasars (e.g.\ \citealt{Miller2009,Miller2011}) with typical quasars, and to compare high and low redshift quasars (e.g.\ \citealt{Vito2019}). While our investigation required a large number of sources to improve the confidence of and explore the X-ray--UV emission-line correlations, we also sought to add much needed data to Sample~B. To generate a subsample of quasars with a very high detection fraction (akin to Sample B) using our data, the reader can simply impose an \mbox{$i$-magnitude} cut of $i \geq 20.2$ to our Sensitive sample. As depicted in Figure \ref{fig:Appdx1} by the dashed black curve, this magnitude cut will remove all but two of the X-ray upper limits (red squares) from our Sensitive sample (green points). After this cut, the total number of quasars with \mbox{X-ray} and SDSS spectral measurements becomes 304 (302 quasars X-ray detected, 2 X-ray upper limits). This subsample contains approximately twice the number of quasars that are in \citet{Gibson2008} Sample B, and has a 99.3\% detection fraction (compared to the 96.3\% detection fraction for our Sensitive sample; see Table \ref{tab:Samples}). We recommend this subsample for projects that require the highest X-ray detection fraction rather than a large sample size.

\section{The Full Sample}\label{sec:appendixB}
Here we present the table containing the X-ray and UV emission line measurements for the quasars in our Full sample. Table \ref{tab:FullCatalog} reports ten of the 95 columns in the full catalog. We report null values as either $-1$ for the quasar spectra in the Full sample that were not fit (or for objects that are not covered by the FIRST), and $-999.0$ for quasars where we attempted to fit the spectrum, but the fitting failed (typically when the \mgii emission line was exiting the spectral window at $z \approx 2.25$). This table is fully available online in machine-readable format. Below, we provide a brief description for each of the columns in Table \ref{tab:FullCatalog}.
\begin{table}\label{tab:FullCatalog}
\begin{adjustbox}{addcode={\begin{minipage}{\width}}{\caption{%
Selected columns from the table containing our Full data set. Null values for the emission-line measurements are represented by $-1$, indicating no fit was attempted, or $-999.0$, meaning the fit was attempted but failed. The number after the underscore in column (7) indicates that more than one quasar was detected in a single observation ID. Descriptions of all columns can be found in Appendix \ref{sec:appendixB}. The entire table is available online in machine-readable format.
}\end{minipage}},rotate=90,center}

\begin{tabular}{r r r c c c c r r r}
\hline
\thead{SDSS Name}  & \thead{RA} & \thead{DEC} & \thead{$z$} & \thead{ObsID} & \thead{$\theta$ }& \thead{${\rm{log}}_{10}(L_{2500})$} & \thead{\daox} & \thead{\civew} & \thead{\mgiiew} \\
\thead{} & \thead{(J2000 deg)} & \thead{(J2000 deg)} &  &  & \thead{(arcmin)} & \thead{($\mathrm{erg\,s^{-1}\,Hz^{-1}}$)} &  & \thead{($\mathrm{\mathring{A}}$)} & \thead{($\mathrm{\mathring{A}}$)} \\
\thead{(1)} &\thead{(2)} & \thead{(3)} & \thead{(4)} & \thead{(7)}  & \thead{(8)} & \thead{(38)} & \thead{(42)} & \thead{(67)} & \thead{(87)} \\
\hline
$022540.57-043825.2$ & 36.4191 & $-$4.6403 & 2.482 & 6864\_1 & 7.91 & 30.7868 & 0.0026 & 37.9655 & $-$999.0 \\
$091945.18+011118.6$ & 139.9383 & 1.1885 & 2.259 & 7056\_1 & 10.6187 & 30.0721 & 0.1273 & $-$1.0 & $-$1.0 \\
$235358.62+293102.5$ & 358.4943 & 29.5174 & 2.4298 & 16441 & 8.7007 & 30.7224 & 0.2496 & $-$1.0 & $-$1.0 \\
$092150.89+293547.0$ & 140.462 & 29.5964 & 2.3697 & 12187\_1 & 7.7575 & 30.3683 & $-$0.1682 & $-$1.0 & $-$1.0 \\
$111118.52+405814.4$ & 167.8272 & 40.9707 & 1.7343 & 15106 & 6.3252 & 30.0221 & 0.095 & $-$1.0 & $-$1.0 \\

\hline

\end{tabular}

\end{adjustbox}
\end{table}

\begin{itemize}

\item[--] Column (1): SDSS Name
\item[--] Column (2): J2000 Right Ascension (J2000 degrees)
\item[--] Column (3): J2000 Declination (J2000 degrees)
\item[--] Column (4): Redshift (see \citealt{Shen2011,Paris2017,Paris2018})
\item[--] Column (5): Galactic column density (cm$^{-2}$; \citealt{Kalberla2005})
\item[--] Column (6): Absolute magnitude (corrected to $z=2$; \citealt{Richards2006})
\item[--] Column (7): \Chandra observation ID (The number after the underscore indicates that more than one quasar was detected in a single observation ID. This value can be disregarded.)
\item[--] Column (8): \Chandra off-axis angle (arcmin)
\item[--] Column (9): Soft-band effective exposure time (seconds)
\item[--] Column (10): Hard-band effective exposure time (seconds)
\item[--] Column (11): Binomial probability of detection (soft band)
\item[--] Column (12): Binomial probability of detection (hard band)
\item[--] Column (13)-(18): Net counts (or 90\% confidence upper limits); upper and lower limits of the soft- and hard-band counts
\item[--] Column (19)--(21): Hardness ratio and lower and upper limits
\item[--] Column (22)--(24): Power-law photon index (dual-band detections) or limit (single-band detection) and upper/lower limits
\item[--] Column (25)--(27): Soft-band count rate (s$^{-1}$); upper and lower limits
\item[--] Column (28)--(30): 0.5--2 keV flux (erg cm$^{-2}$  s$^{-1}$); lower and upper limit
\item[--] Column (31)--(33): Observed flux density at rest-frame 2 keV \mbox{(erg cm$^{-2}$  s$^{-1}$ Hz$^{-1}$)}; lower and upper limit
\item[--] Column (34): Rest-frame 2--10 keV luminosity (erg s$^{-1}$); lower and upper limit
\item[--] Column (37): Observed flux density at rest-frame 2500 \angstrom \mbox{(erg cm$^{-2}$  s$^{-1}$ Hz$^{-1}$)}
\item[--] Column (38): Rest-frame 2500 \angstrom monochromatic luminosity (erg s$^{-1}$ Hz$^{-1}$)
\item[--] Column (39)--(41): Observed \aox; lower and upper limit
\item[--] Column (42): Difference between observed \aox and the expectation from the \aox--$L_{2500}$ relation (this work)
\item[--] Column (43): Difference between observed \aox and the expectation from the \aox--$L_{2500}$ relation in \citet{Just2007}
\item[--] Column (44)--(45): \daox lower limit and upper limit (this work)
\item[--] Column (46): Flag if the target overlapped with the edge of the \Chandra CCD (1 for yes, 0 for no)
\item[--] Column (47): Date of the \Chandra observation
\item[--] Column (48): \Chandra observation cycle
\item[--] Column (49): Observed more than once with \Chandra (1 for yes, 0 for no)
\item[--] Column (50)--(52): SDSS spectroscopic plate ID, MJD, \mbox{fiber ID}
\item[--] Column (53)--(57): De-reddend $u,g,r,i,z$ band apparent magnitude from SDSS (using \citealt{Schlafly2011} dust map)
\item[--] Column (58): SDSS catalog in which the target was first published (DR7, 12, 14)
\item[--] Column (59): Flag for detected \civ broad absorption lines (0= no line measurement, 1= BAL, 2= mini-BAL)
\item[--] Column (60): Flag indicating if the source was detected in FIRST ($-1$ = target not covered by FIRST, $0$ = FIRST non-detection, $1$ = FIRST detected)
\item[--] Column (61): SDSS relative $g - i$ color \citep{Richards2011}
\item[--] Column (62): Sensitive sample flag indicating if the target passed the exposure time -- off-axis angle cut (1 = passed, 0 = did not pass)
\item[--] Column (63): Measured signal-to-noise of the line-free continuum regions
\item[--] Column (64): Logarithm of observed flux density at rest-frame 2500 \angstrom \mbox{(erg cm$^{-2}$  s$^{-1}$ Hz$^{-1}$)} measured directly from the spectrum 
\item[--] Column (65): Rest-frame equivalent width (\angstrom) of the \civ broad emission line measured directly from the spectrum 
\item[--] Column (66): Measurement error of rest-frame equivalent width (\angstrom) of the \civ broad emission line measured directly from the spectrum
\item[--] Column (67): Rest-frame equivalent width (\angstrom) of the \civ broad emission line measured from the Gaussian models
\item[--] Column (68): Measurement error of rest-frame equivalent width (\angstrom) of the \civ broad emission line measured from the Gaussian models
\item[--] Column (69): Line dispersion (km s$^{-1}$) of the \civ broad emission line
\item[--] Column (70): Measured rest-frame peak of the \civ broad emission line
\item[--] Column (71): Uncertainty in the measured rest-frame peak of the \civ broad emission line
\item[--] Column (72): Full-width at half-maximum of the modeled \civ broad emission line
\item[--] Column (73): Uncertainty in full-width at half-maximum of the modeled \civ broad emission line
\item[--] Column (74): Median signal-to-noise ratio per pixel for the \civ broad emission line
\item[--] Column (75)--(84): Same as Columns (65)--(74) for the \ion{C}{iii]} emission line
\item[--] Column (85)--(94): Same as Columns (65)--(74) for the \ion{Mg}{ii} emission line
\item[--] Column (95): Radio-loudness parameter (or upper limit if FIRST\_flag $=$ 0)

\end{itemize}




\bibliographystyle{mnras}
\bibliography{DR7_12_WLQ_correlation_v1} 



\bsp	
\label{lastpage}
\end{document}